\documentclass[twocolumn]{aastex63}
\usepackage{float}
\usepackage{amsmath}
\usepackage{amssymb}
\usepackage{latexsym}
\usepackage{textcomp}
\usepackage{graphicx}
\usepackage{subfigure}
\usepackage{epsfig}
\usepackage{natbib}

\newcommand{\hi}{\textrm{H}~\textsc{i}}
\newcommand{\ha}{\textrm{H}\ensuremath{\alpha}}
\newcommand{\hb}{\textrm{H}\ensuremath{\beta}}

\newcommand{\hei}{\textrm{He}\,\textsc{i}}
\newcommand{\nii}{[\textrm{N}~\textsc{ii}]}

\newcommand{\oiii}{[\textrm{O}~\textsc{iii}]}
\newcommand{\hii}{\textrm{H}~\textsc{ii}}
\newcommand{\sii}{[\textrm{S}~\textsc{ii}]}
\newcommand{\oi}{[\textrm{O}~\textsc{i}]}

\newcommand{\oiiilam}{[\textrm{O}~\textsc{iii}]~\ensuremath{\lambda5007}}
\newcommand{\niilam}{[\textrm{N}~\textsc{ii}]~\ensuremath{\lambda6584}} 
\newcommand{\niilamboth}{[\textrm{N}\,\textsc{ii}]\,\ensuremath{\lambda\lambda6548,6584}}

\newcommand{\oilam}{[\textrm{O}~\textsc{i}]~\ensuremath{\lambda6300}}
\newcommand{\siilam}{[\textrm{S}~\textsc{ii}]~\ensuremath{\lambda\lambda6717,6731}}

\newcommand{\oiiilamboth}{[\textrm{O}\,\textsc{iii}]\,\ensuremath{\lambda\lambda4959,5007}}

\newcommand{\htwo}{\textrm{H}~\ensuremath{\lambda2.122\,\mu m}} 
\newcommand{\feiilam}{[\textrm{Fe}~\textsc{ii}]~\ensuremath{\lambda1.644\,\mu m}} 
\newcommand{\brgam}{\textrm{Br}\ensuremath{\gamma}}
\newcommand{\ergs}{\textrm{erg\,s$^{-1}$}}
\newcommand{\kms}{\textrm{km\,s$^{-1}$}}

\newcommand{\Msun}{$M_{\sun}$}

\shorttitle{AGN-Host Connection: Role of Galaxy Substructure}
\shortauthors{Juneau et al.}

\begin{document}

\title{The Black Hole-Galaxy Connection: \\ Interplay between Feedback, Obscuration, and Host Galaxy Substructure}

\author[0000-0002-0000-2394]{St\'{e}phanie Juneau}
\affiliation{NSF's NOIRLab, 950 N. Cherry Ave., Tucson, AZ 85719, USA}
\affiliation{CEA-Saclay, DSM/IRFU/SAp, 91191 Gif-sur-Yvette, France}
\email{stephanie.juneau@noirlab.edu}

\author[0000-0003-4700-663X]{Andy D. Goulding}
\affiliation{Department of Astrophysical Sciences, Princeton University, Ivy Lane, Princeton, NJ 08544, USA}

\author[0000-0003-4417-5374]{Julie Banfield}
\affiliation{Research School of Astronomy and Astrophysics, Australian National University, Canberra ACT 2611, Australia}

\author[0000-0002-4622-4240]{Stefano Bianchi}
\affiliation{Dipartimento di Matematica e Fisica, Universit\`a degli Studi Roma Tre, via della Vasca Navale 84, 00146 Roma, Italy}

\author[0000-0003-3343-6284]{Pierre-Alain Duc}
\affiliation{Observatoire astronomique de Strasbourg, Universit\'e de Strasbourg, CNRS, UMR 7550, 11 rue de l'Universit\'e, F-67000 Strasbourg, France}
\affiliation{CEA-Saclay, DSM/IRFU/SAp, 91191 Gif-sur-Yvette, France}

\author[0000-0002-0757-9559]{I-Ting Ho}
\affiliation{Max Planck Institute for Astronomy, K\"onigstuhl 17, 69117 Heidelberg, Germany}

\author[0000-0003-0922-4986]{Michael A. Dopita}
\altaffiliation{Deceased}
\affiliation{Research School of Astronomy and Astrophysics, Australian National University, Cotter Rd., Weston ACT 2611, Australia}

\author[0000-0003-1585-9486]{Julia Scharw\"achter}
\affiliation{Gemini Observatory/NSF's National Optical-Infrared Astronomy Research Laboratory, 670 N. A'ohoku Place, Hilo, HI, 96720, USA}

\author[0000-0002-8686-8737]{Franz E. Bauer}
\affiliation{Instituto de Astrof{\'{\i}}sica and Centro de Astroingenier{\'{\i}}a, Facultad de F{\'{i}}sica, Pontificia Universidad Cat{\'{o}}lica de Chile, Casilla 306, Santiago 22, Chile} 
\affiliation{Millennium Institute of Astrophysics (MAS), Nuncio Monse{\~{n}}or S{\'{o}}tero Sanz 100, Providencia, Santiago, Chile} 
\affiliation{Space Science Institute, 4750 Walnut Street, Suite 205, Boulder, Colorado 80301} 

\author[0000-0002-9768-0246]{Brent Groves}
\affiliation{International Centre for Radio Astronomy Research, The University of Western Australia, 35 Stirling Hw, 6009 Crawley, WA, Australia}
\affiliation{Research School of Astronomy and Astrophysics, Australian National University, Cotter Rd., Weston ACT 2611, Australia}

\author[0000-0002-5896-6313]{David M. Alexander}
\affiliation{Centre for Extragalactic Astronomy, Department of Physics, Durham University, South Road, Durham DH1 3LE, UK}

\author[0000-0002-3324-4824]{Rebecca L. Davies}
\affiliation{Max-Planck-Institut f\"ur Extraterrestrische Physik, Giessenbachstrasse, D-85748 Garching, Germany}
\affiliation{Research School of Astronomy and Astrophysics, Australian National University, Cotter Rd., Weston ACT 2611, Australia}

\author[0000-0002-7631-647X]{David Elbaz}
\affiliation{CEA-Saclay, DSM/IRFU/SAp, 91191 Gif-sur-Yvette, France}

\author[0000-0002-9061-5409]{Emily Freeland}
\affiliation{Department of Astronomy, The Oskar Klein Center, Stockholm University, Albanova, SE 10691 Stockholm, Sweden}

\author{Elise Hampton}
\affiliation{Research School of Astronomy and Astrophysics, Australian National University, Cotter Rd., Weston ACT 2611, Australia}

\author[0000-0001-8152-3943]{Lisa J. Kewley}
\affiliation{Research School of Astronomy and Astrophysics, Australian National University, Cotter Rd., Weston ACT 2611, Australia}

\author[0000-0002-7052-6900]{Robert Nikutta}
\affiliation{NSF's NOIRLab, 950 N. Cherry Ave., Tucson, AZ 85719, USA}

\author[0000-0002-4984-9641]{Prajval Shastri}
\affiliation{Indian Institute of Astrophysics, Sarjapur Road, Bengaluru 560034, India}

\author{Xinwen Shu}
\affiliation{Department of Physics, Anhui Normal University, Wuhu, Anhui, 241000, China}
\affiliation{CEA-Saclay, DSM/IRFU/SAp, 91191 Gif-sur-Yvette, France}

\author[0000-0002-9665-2788]{Fr\'ed\'eric P. A. Vogt}
\affiliation{Federal Office of Meteorology and Climatology MeteoSwiss, Chemin de l'A\'erologie 1, 1530 Payerne, Switzerland}
\affiliation{Research School of Astronomy and Astrophysics, Australian National University, Cotter Rd., Weston ACT 2611, Australia}

\author{Tao Wang}
\affiliation{Key Laboratory of Modern Astronomy and Astrophysics in Ministry of Education, School of Astronomy \& Space Science, Nanjing University, Nanjing, 210093, PR China}
\affiliation{CEA-Saclay, DSM/IRFU/SAp, 91191 Gif-sur-Yvette, France}

\author[0000-0003-4264-3509]{O. Ivy Wong}
\affiliation{CSIRO Space \& Astronomy, PO Box 1130, Bentley, WA 6102, Australia, WA 6009, Australia}
\affiliation{International Centre for Radio Astronomy Research, The University of Western Australia, 35 Stirling Hw, 6009 Crawley, WA, Australia}

\author[0000-0002-8055-5465]{Jong-Hak Woo}
\affiliation{Astronomy Program, Department of Physics and Astronomy, Seoul National University, Seoul 08826, Republic of Korea}

\begin{abstract}

There is growing evidence for physical influence between supermassive black holes and their host galaxies. We present a case study of nearby galaxy NGC~7582, for which we find evidence that galactic substructure plays an important role in affecting the collimation of ionized outflows as well as contributing to the heavy active galactic nucleus (AGN) obscuration. This result contrasts with a simple, small-scale AGN torus model, according to which AGN wind collimation may take place inside the torus itself, at subparsec scale. 
Using 3D spectroscopy with the MUSE instrument, we probe the kinematics of the stellar and ionized gas components as well as the ionization state of the gas from a combination of emission line ratios.  
We report for the first time a kinematically distinct core (KDC) in NGC~7582, on a scale of $\sim$600~pc. This KDC coincides spatially with dust lanes and starbursting complexes previously observed. We interpret it as a circumnuclear ring of stars and dusty, gas-rich material. 
We obtain a clear view of the outflowing cones over kpc scales, and demonstrate that they are predominantly photoionized by the central engine. We detect the back cone (behind the galaxy), and confirm previous results of a large nuclear obscuration of both the stellar continuum and \hii\ regions. While we tentatively associate the presence of the KDC to a large-scale bar and/or a minor galaxy merger, 
we stress the importance of gaining a better understanding of the role of galaxy substructure in controlling the fueling, feedback and obscuration of AGN.

\end{abstract}

\section{Introduction}

It has become increasingly clear that a full picture of galaxy formation and evolution requires
an understanding of the interplay between supermassive black holes (SMBHs) and their host galaxies. 
Yet, there are several open questions regarding their underlying physical connections. 
SMBHs reside in the heart of most $-$ if not all $-$ massive galaxies \citep[e.g.,][]{mag98,geb00,fer00}.
When strongly accreting as Active Galactic Nuclei (AGN), they can inject radiative and/or
mechanical energy into their surrounding medium thereby ionizing, heating and/or displacing
it \citep[e.g.,][]{som08,fab12,cie18}. By impacting the gas reservoirs of galaxies, such AGN feedback
could in turn regulate the surrounding star formation. This process is widely implemented in
cosmological simulations as a way to suppress star formation and avoid over-producing the
number of massive, star-forming galaxies relative to observations \citep[e.g.,][]{cro06,bow06,dub13,hir14}.

Observations and simulations have shown a range of results, from AGN feedback having negative (i.e. suppressing) impacts on star formation inferred from the presence of outflows \citep[e.g.,][]{fer10,rup11,bie17,rup17} to positive (i.e., enhancing) impacts from mechanisms such as gas compression \citep[e.g.,][]{gai11,bie16,muk18b}. On the scale of individual galaxies, some studies suggest co-existing negative and positive feedback within a given galaxy \citep[e.g.][]{cre15,shi19}, and numerical simulations predict that the structure of the interstellar medium of gas-rich disk galaxies can play a role in AGN fueling and obscuration \citep{bou11,bou12} as well as in affecting AGN feedback \citep{gai11,gab14,roo15}. The surge of integral field spectroscopy observations of AGN and their host galaxies has generated new opportunities to spatially map and disentangle the gas ionization state and the kinematics of the ionized gas and stellar content \citep[e.g.,][]{mue11,min19,hus19}, as well as possible multi-phase or molecular gas components \citep[e.g.,][]{shi19,shim19,fer20}. Such studies allow us to build a more comprehensive picture of the interplay between AGN feedback and their hosts.

Besides AGN feedback, additional clues on the SMBH-galaxy connection can be obtained from constraints on AGN obscuration, which can also occur on a range of physical scales \citep[see review by][]{ram17}. The most basic AGN unification model predicts that the obscuring medium is a small (parsec scale) nuclear torus surrounding the active black hole (BH), and the degree of obscuration simply depends on the viewing angle \citep{ant93,urr95}. However, there is compelling evidence for AGN obscuration being closely related to their host galaxies. For example, previous work has revealed different hosts and dust structures between Seyfert 1s and 2s down to fine angular scales, strongly disfavoring the conventional torus orientation picture \citep[e.g.,][]{mal98,pri21}. Furthermore, spatially resolved mid-IR or hard X-ray studies of nearby heavily absorbed systems found that the obscuration is fully consistent with moderate-to-large scale dust lanes and features with a range of Hydrogen column densities \citep{bia07,are14,bau15}, and with AGN ionization cone collimation taking place at distances well beyond the inner torus \citep{pri14,mez16}. 
Additionally, the observed relationship between the 9.7$\mu$m Silicate absorption depth and the inclination of the host galaxies of heavily absorbed AGN support large-scale obscuration \citep{gou12}, and an intriguing trend for intermediate redshift star-forming galaxies points to an increasing fraction of heavily absorbed AGN with increasing galaxy-wide specific star formation rate \citep[=SFR/stellar mass, a tracer of galaxy gas fractions;][]{jun13}. The latter suggests either a direct role of host galaxies gas in the absorption of X-rays from the AGN, or a true physical link between the small-scale torus and the multi-scale interstellar medium. 
Some studies find cases with high X-ray absorption (i.e., Compton-thick regime) which are consistent with small-scale {\it torus} absorption \citep[e.g.,][]{bal14,marko14,ric14,kos17}, or evidence for both nuclear and possible host galaxy contributions, where the latter might be galaxy mass- and redshift-dependent \citep{bri14,buc15,buc17}.
Therefore, assessing AGN feedback, AGN obscuration, and the possible interplay with host galaxy properties and substructure can shed light on the SMBH-galaxy connection.
  
In this paper, we investigate these questions using a case study of the Compton-thick AGN host \object[NGC 7582]{NGC~7582}, for which a few lines of evidence point toward an interesting connection between the galaxy substructure and the central AGN, including indications that AGN collimation and/or obscuration may reach physical scales larger than the putative torus from the AGN unification model. By taking advantage of the combined large field-of-view, fine spatial sampling, and high sensitivity of the Multi Unit Spectroscopic Explorer \citep[MUSE;][]{bac10} on the Very Large Telescope (VLT), we paint an overall picture of the stellar kinematics, gas kinematics and gas excitation properties, and revisit the obscuration at a range of physical scales. The target and observations are described in Sections~\ref{sec:sample} and \ref{sec:obs}, followed by the method used to fit the spectra and extract physical parameters (Section~\ref{sec:method}), before we report our results in Section~\ref{sec:result}. 
Lastly, our main findings are discussed and summarized in Sections~\ref{sec:discu} and \ref{summ}.  Throughout this paper, we adopt a 
flat $\Lambda$CDM cosmology  ($\Omega_m = 0.3$, $\Omega_{\Lambda} = 0.7$, and $h = 0.7$) and a \citet{cha03} initial mass function.

\section{The Target Galaxy}\label{sec:sample}

NGC 7582 is a nearby, inclined, barred spiral (SBab; axis ratio of 0.42)\footnote{From the NASA Extragalactic Database: https://ned.ipac.caltech.edu}. Its central source is a 
Compton-thick AGN usually classified as a Type 2 Seyfert from its optical spectrum. 
However, there is reported variability in both the X-ray absorption \citep{pic07,bia09,riv15,lah20}, and 
optical classification, which we describe below. 
The central black hole is estimated to have a mass of $\sim5.5\times10^7$~\Msun\ \citep{wol06}, and it has a boxy peanut-shaped bulge identified from near-infrared imaging \citep{qui97}.

NGC~7582 was previously known to have extended narrow-line region emission from \oiiilam\ (hereafter \oiii) narrow band imaging \citep{mor85,sto91,rif09} and emission-line maps from the Siding Spring Southern Seyfert Spectroscopic Snapshot Survey \citep[S7;][]{dop15b,tho17}. The S7 integral field spectroscopy datacube was obtained with the Wide Field Spectrograph \citep{dop07}. Their analysis confirmed the extended emission line regions along the front cone, but also revealed an optical view of the counter-cone on the far side of the galaxy \citep{dav16} which was originally found from extended soft X-ray emission \citep{bia07}. \citet{davr20} computed an ionized gas mass outflow rate of 0.007~\Msun\,yr$^{-1}$ based on spectra from the VLT/Xshooter instrument covering the inner $\sim$300~pc.

While the optical spectrum often only shows narrow lines, broad lines are present in the infrared regime \citep[e.g., broad Br$\gamma$;][]{sos01,davi05}, which led \citet{ver06} to classify it as S1i on their system (where $i$ stands for infrared). A previous finding of a broad \ha\ component was interpreted as possibly due to supernovae contributions by \citet{are99} though the spatial resolution was insufficient to firmly distinguish between the AGN and SN scenarios. 
Recently, \citet{ricc18} reported a broad \ha\ component by isolating the central region using IFU observations, which could be consistent with a true Type 1 AGN behind a dust screen, though the authors' interpretations favor either a partial view of the broad line region through a clumpy torus or a reflected component from the inner part of the ionization cone. 
However, the infrared coronal line [Si VII] is characterized by an isotropic morphology \citep{pri14} in contrast with the cone shape \oiii\ emission, thus supporting evidence for a true type 1 AGN classification.  Other signatures of obscuration include a significant Silicate (Si 9.7$\mu$m) absorption feature associated with foreground host galaxy material \citep{gou12}, and X-ray spectral analysis indicating the presence of at least two absorbers on different spatial scales \citep{bia07,pic07}, and a significant Fe K$\alpha$ equivalent width \citep[EW$\sim 269-639$~eV depending on the X-ray spectrum model;][]{brig11a}. The variability of observed AGN signatures coupled with various interpretations have contributed to NGC 7582's reputation as an {\it interesting puzzle} \citep[their \S 5.31]{sos01}. Taken together, these pieces of evidence point toward intriguing possible connections between the host galaxy, its internal structure and the central AGN in NGC~7582. 

\begin{figure*}[hbt]
\includegraphics[width=1\textwidth,bb=0 0 1139 650]{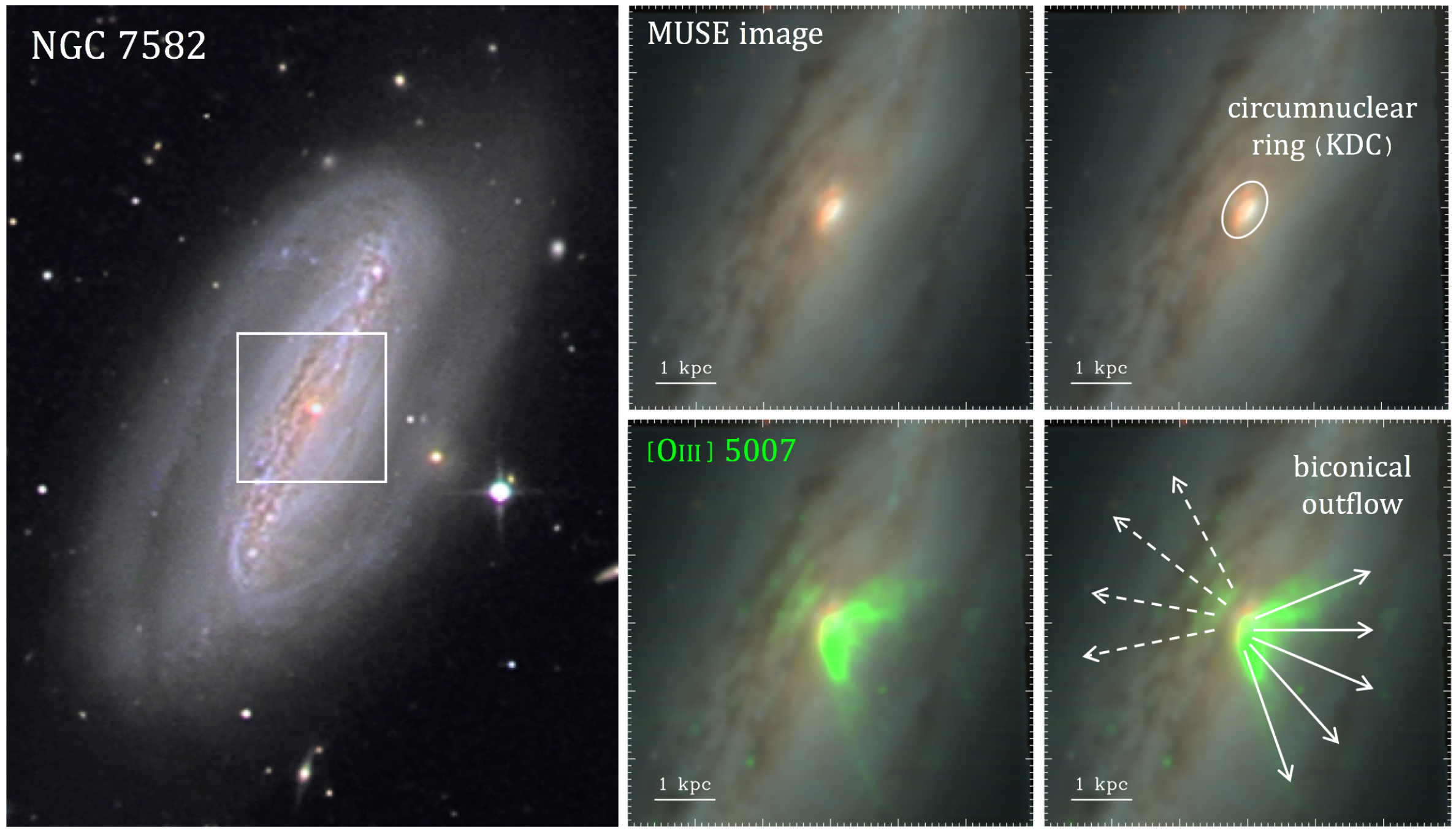}
\caption{Color images of NGC~7582. (Left) LRGB composite image showing the galaxy-scale view. The MUSE field-of-view is marked with the white square, and covers $1\arcmin \times 1\arcmin$. Image credit: S. Binnewies \& J. P\"{o}psel from Capella Observatory. (Top Center) Reconstructed RGB (Red-Green-Blue) color image from the MUSE spectral cube, as described in Section~\ref{sec:dust}. We clearly see the dust lanes against the stellar continuum as well as the notably dusty circumnuclear region. (Top Right) Same as the Top Center with additional markings showing the dusty circumnuclear ring. (Bottom Center) Reconstructed RGB image with the addition of the \oiii\ emission, which mostly traces the ionized, biconical outflows, in the green (G) channel. (Bottom Right) Same as the Bottom Center panel but with arrows that illustrate the front cone (solid lines) and back cone (dashed lines). The scale bar measures 9\arcsec, which corresponds to 1~kpc at the distance to NGC~7582 of 22.7~Mpc. The typical seeing of the observations, with FWHM$\sim$1\arcsec, corresponds to a physical size of $\sim$110~pc and is represented by one minor tick mark on the MUSE images, which encompass the central $60\arcsec \times 60\arcsec$ of the MUSE field-of-view. All MUSE maps were smoothed with a Gaussian kernel with FWHM~$\sim1\farcs4$. On all panels, North is up, East is to the left.
}\label{fig:labels}
\end{figure*}

\begin{deluxetable}{lcccccc}
\tablecolumns{7}
\tablewidth{0pc}
\tablecaption{Summary of Target Properties\label{tab:samples}}
\tablehead{
   \colhead{Galaxy}  &  \colhead{$D_L$} &  \colhead{Morph.}  &  \colhead{b/a}  & \colhead{$L_X$}  & \colhead{$N_H$}         &  \colhead{S9.7} \\
   \colhead{ }            &  \colhead{Mpc}        &  \colhead{ }             &   \colhead{ }  &  \colhead{\ergs}  & \colhead{${\rm 10^{24}\,cm^{-2}}$} &  \colhead{ } 
}
\startdata
NGC7582   &   22.7    &  SBab    &     0.42   &  42.61    &  1.6  &  0.78 \\
\enddata
\tablenotetext{}{Columns: (1) Galaxy name; (2) Luminosity distance in Mpc corrected for non-cosmological flow; (3) Galaxy morphological type; (4) Galaxy axis ratio; (5) Logarithm of X-ray luminosity at 2$-$10~keV corrected for absorption; (4) Hydrogen column density; (5) Silicate absorption strength \citep[average values from][who reported X-ray information from \citet{tur00,mar05}]{gou12}.}
\end{deluxetable}

\section{Observations and Data Reduction}\label{sec:obs}

To reveal physical clues about the multi-scale BH-galaxy relation, we used the MUSE instrument with a 1~square arc min field-of-view, corresponding to 8~kpc on a side at the distance to NGC~7582. Figure~\ref{fig:labels} shows an overview of the target galaxy with an overlay of the MUSE field-of-view as well as reconstructed color images obtained from the MUSE datacube without and with the inclusion of the \oiii\ line emission, which we added in the green channel of the color image. We present these images as a reference with labels of the important components that will be studied and discussed throughout the rest of this paper. The method used to construct color images from the MUSE datacube is described in Section~\ref{sec:dust}. 

NGC~7582 was observed on 2015-08-07 with a seeing around 1\farcs0 as part of ESO program 095.A-0934 (PI Juneau). We combined four exposures of 10 minutes, adding to a total of 40 minutes on-source. We changed the rotation angle by 90 degrees between each object exposure (O), and acquired sky frames (S) of 60s following the sequence O-O-S-O-O-S. Rotating the instrument between exposures was recommended by ESO in order to help correct for patterns of the slicers and channels when co-adding the exposures, and therefore obatin more uniform noise properties.

The data were processed with the ESO Reflex \citep[v1.0.5]{freu13} implementation of the MUSE data reduction pipeline \citep{wei15}. The reduced data cube has spaxels of $0\farcs2 \times 0\farcs2$ on the sky with wavelength spacing of 1.25\,\AA, which corresponds to $\sim55-75$~\kms over the wavelength range of interest. The MUSE spectral resolution varies from 1750 at 4650\,\AA\ to 3750 at 9300\,\AA. The full spectral range covers 4750$-$9350\,\AA\ though we fit the spectra up to 8900\AA, and focus the emission-line analysis at $\lambda<$6800\,\AA. 
A stat cube of the same dimension contains the variance, which is helpful to check overlap with strong sky lines, and to mask them during spectral fitting (Section~\ref{sec:fit}).

Our MUSE program supplements previous observations by providing improved sensitivity and a larger field-of-view, and in the case of S7, higher spatial resolution as well. Indeed, one can see a high level of details with very sharp edges defining the front ionization cone shown in green with solid arrows in Figure~\ref{fig:labels}, while also finding clear indications of the counter (back) cone between the dust lanes (dashed arrows). 

\section{Method}\label{sec:method}

\subsection{Spectral fitting}\label{sec:fit}

The MUSE datacube is fitted for both stellar continuum and emission lines using LaZy-IFU \citep[LZIFU;][]{ho16}. 
This IDL\footnote{Interactive Data Language: www.l3harrisgeospatial.com/Software-Technology/IDL}-based code 
is publicly available\footnote{https://github.com/hoiting/LZIFU/}. 
LZIFU models the continuum with the penalized pixel-fitting routine \citep[PPXF;][]{capp04}. 
During continuum fitting, emission lines and sky lines are masked. Given the spectral range of 
MUSE ($\lambda_{obs}=4750-9350$\,\AA) and the redshift of our target \citep[$z=0.0058$;][]{reu03}, the main absorption lines 
that do not overlap with an emission line mask include Na D and the Ca triplet. We restricted the fit to $\lambda<8900$\,\AA, 
and adopted the MILES simple stellar population libraries \citep{san06,fal11}. Before fitting, the software aligns the stellar 
models and the observed spectra in terms of their wavelength coverage, spectral resolution, and channel width. The continuum 
fitting step then uses linear combinations of stellar population templates to solve simultaneously for stellar 
velocity, stellar velocity dispersion, and stellar reddening \citep{ho16}.

\begin{figure*}[ht]
\epsscale{1.15} \plotone{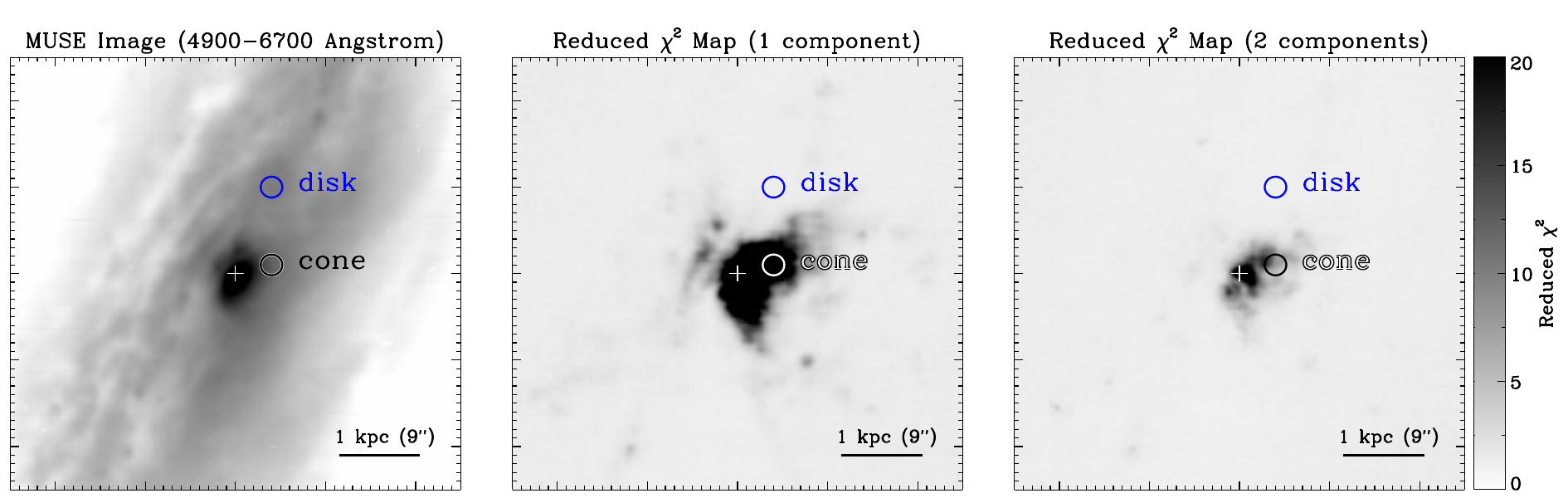}
\caption{
   (Left) MUSE synthetic image from collapsing the spectral cube over rest wavelength $4900-6700$\,\AA. 
   The scale bar measures 9\arcsec, which corresponds to 1~kpc at the distance of NGC~7582. The 2\farcs4 diameter circular apertures 
   used to extract example spectra and spectral fits over a region on the disk, and over 
   a region on the outflowing cones are plotted with circles, and labeled accordingly.
   (Center) Reduced $\chi^2$ map resulting from the fit with 1 component emission lines shown on the same scale as the 
   MUSE image from the left-hand panel. 
   (Right) Reduced $\chi^2$ map resulting from the fit with 2 component emission lines shown on the same scale as the 
   MUSE image from the left-hand panel. The reduced $\chi^2$ maps range from 0 (white) up to 20 (in black) as shown on the color bar. 
   Each panel spans the central $50\arcsec \times 50\arcsec$ of the MUSE field-of-view, with major tick marks every 10\arcsec. 
}\label{fig:chisq}
\end{figure*}

Continuum subtracted spectra are then used by LZIFU to fit an input list of emission lines simultaneously. The fitting procedure assumes 
that the continuum was properly subtracted and that the lines can be described by one, two, or three Gaussian profiles 
where each Gaussian component has common kinematic properties (velocity offset and velocity dispersion) for all the lines. 
We performed a first fit over the datacube assuming a single Gaussian component for the emission lines, and a second fit assuming 
two Gaussian components. The fitting procedure minimizes the reduced chi-squared ($\chi^2$) by fitting all emission lines simultaneously with the Levenberg-Marquardt least-square method MPFIT \citep{mar09}. Output quantities from the emission line fitting include gas kinematics (velocity and velocity dispersion) as well as emission line fluxes and uncertainties for the single (or double) Gaussian components. We compare the goodness-of-fit from the fits with one component or two component emission lines with the reduced $\chi^2$ maps in Figure~\ref{fig:chisq}. The left-hand panel shows a synthetic MUSE image obtained by collapsing the spectra cube over rest wavelengths $4900-6700$\,\AA\ with spectral extraction apertures selected to define one region over the disk (avoiding dust lanes), and another region over the \oiii\ cone shown in Figure~\ref{fig:labels}. The middle panel displays the reduced $\chi^2$ map resulting from the single-component fit, which clearly shows very elevated values coinciding with the \oiii\ cone (black color corresponds to reduced $\chi^2>20$). The right-hand panel displays the reduced $\chi^2$ map from the double-component fit, which is much improved relative to that from the single-component fit although there remain areas with comparatively high $\chi^2$ values over the central region. 

\begin{figure*}[hb]
\epsscale{.9} \plotone{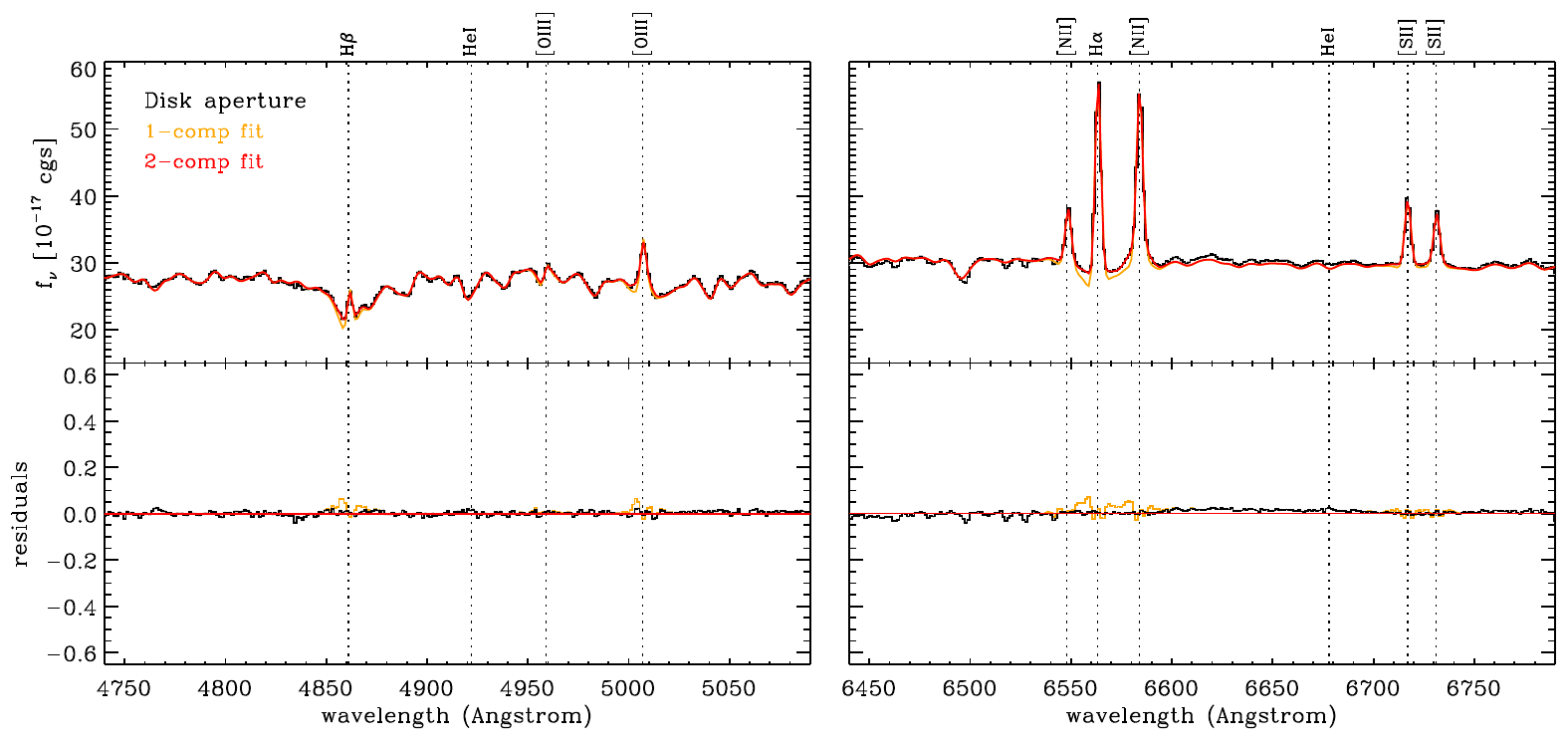}
\epsscale{.9} \plotone{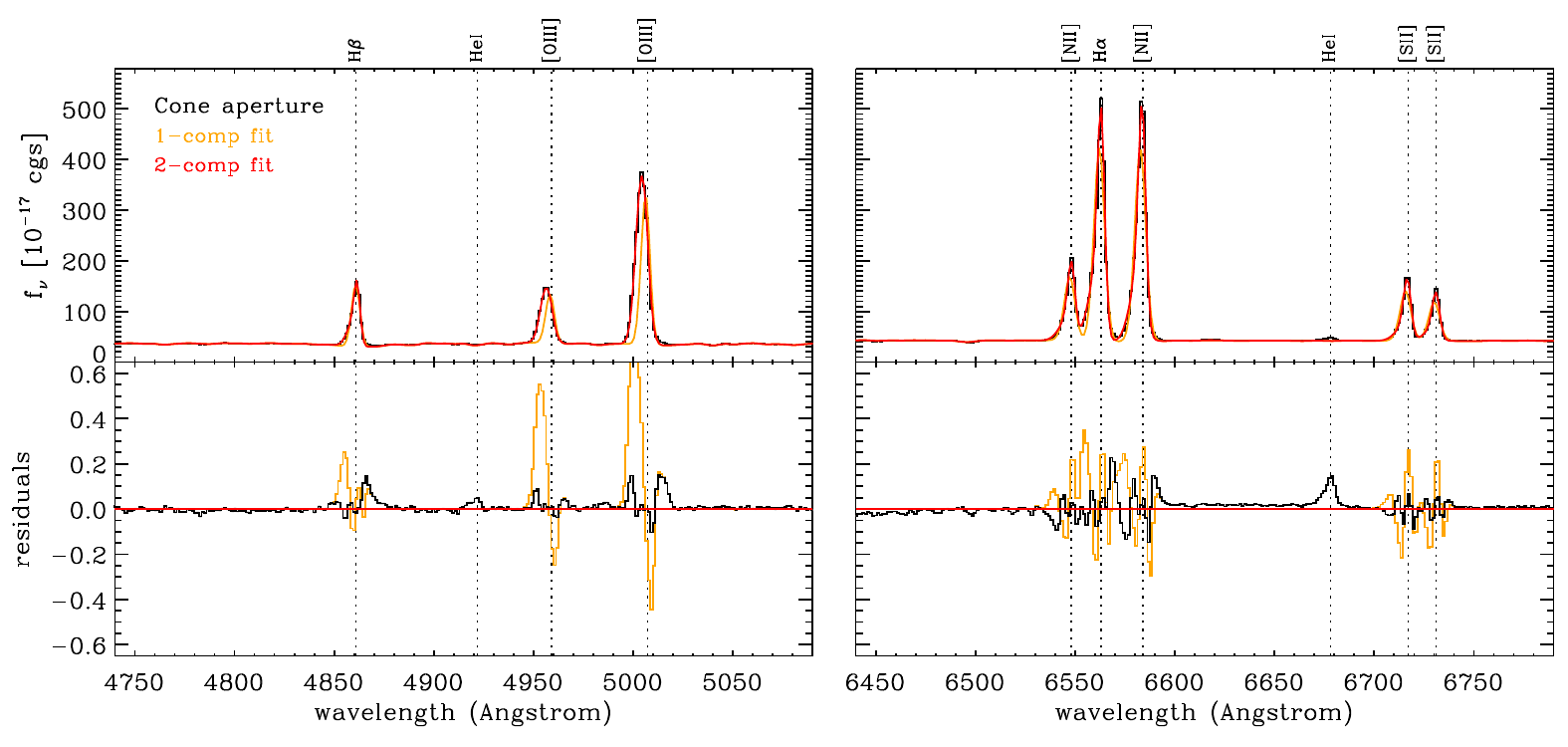}
\caption{
   MUSE spectrum (black) and LZIFU fits (orange and red) extracted from the \emph{disk} aperture (top panels), 
   and \emph{cone} aperture (bottom panels) drawn in Figure~\ref{fig:chisq}. Each set of panels includes the spectra 
   around the \hb\ and \oiii\ doublet region (left-hand side), and around the \ha, \nii\ doublet and \sii\ doublet 
   region (right-hand side). In all cases, the bottom panels show the relative residuals for both the single-component 
   fit (orange), and two-component fit (black). In the case of the disk aperture, the two fits are more comparable 
   with residuals under 5\%. In the case of the cone aperture, there is a more striking difference, where the 
   single-component fit clearly does not represent the data well, while the two-component fit is better adapted. 
   Note that the \hei\ lines were not included in the fit as they are mostly absent from the datacube.
}\label{fig:fitspec}
\end{figure*}

The spectra and LZIFU spectral fits extracted at the location of the apertures drawn on Figure~\ref{fig:chisq} are displayed in Figure~\ref{fig:fitspec}. We focus on two spectral regions: the first one around \hb\ and \oiiilamboth\ (left-hand side), and the second one around \ha, \niilamboth, and \siilam\ (right-hand side). The top set of panels include the results from the aperture located on the disk. The observed spectrum (black line) is fairly well represented by both the single-component fit (orange) and two-component fit (red). In both cases, the residuals are within $<$5\% (lower panels; black from the double-component fit, orange for the single-component fit). The bottom set of panels include the results from the aperture located on the cone. In contrast to the disk case, the one-component fit does not represent the data adequately, with strong residuals reaching $\pm$40\%. The two-component fit is a better representation with residuals largely $<10-20$\% (black line) over the emission lines. We note that the \hei\ lines were not included in the fit as they are mostly absent from the datacube (besides the regions ionized by the AGN; Section~\ref{sec:ratio}). This explains why the \hei\ lines are visible in the residuals over the cone region.

In this work, we are generally interested in the following strong-lines: \hb, \oiii, \oilam\ (hereafter \oi), \ha, \niilam\ (hereafter \nii), and \siilam\ (hereafter \sii). Preliminary analysis revealed that the most dominant gas kinematic components arise from separating the motion of the galactic disk (for which \ha\ is the dominant/strongest line), and that of the ionization cone (for which \oiii\ is the dominant line). We thus compare the resulting fits using either one or two Gaussian components and we identify spaxels where two components are detected with signal-to-noise S/N$>$3 for \ha\ or for \oiii. Over the full MUSE field-of-view (102708 spaxels), we find two components in 54658 spaxels ($\sim53\%$), while 24\% have one component, and 23\% lack a S/N$>3$ detection in both lines. The latter tend to be spread over the regions away from the galaxy major axis and away from the cones, where there is weak to no detectable signal in the datacube. 
When two components are detected with $S/N>3$ for \ha\ and/or \oiii, we assign the component with velocity closest to systemic velocity (assumed to be the stellar velocity derived during the continuum fitting for a given spaxel) to be Component 1 (c1) and the other to be Component 2 (c2). Spaxels that only have component c1 are masked on the raw c2 velocity map. Spaxels that are undetected ($S/N<3$) in both \ha\ and \oiii\ are masked in the raw c1 velocity map. Smoothing with 2D Gaussian filter with a 3-spaxel width interpolates over the masked values, and produces the final maps shown in Figure~\ref{fig:vel}(b) and (c) for components c1 and c2, respectively.

\subsection{Physical Parameters}

We used the LZIFU fitting results to derive the following physical properties:
\begin{enumerate}

\item[a)] Stellar velocity and velocity dispersion: they are obtained respectively from the best-fit absorption bands and line positions relative to the expected wavelengths given the galaxy redshift, and from the width of the lines.

\item[b)] Gas velocity and velocity dispersion: they are obtained respectively from the emission line positions relative to systemic, and from the width of the lines. 

\item[c)] Dust attenuation: we obtain the stellar reddening, E(B-V), from stellar continuum fitting and the dust attenuation to the gas component from the Balmer Decrement (\ha/\hb).

\item[d)] Gas excitation: the BPT \citep{bpt} and VO87 \citep{vei87} emission line ratio diagnostic diagrams are used to constrain the source of ionization for each spaxel (AGN, SF, shocks). We use the total line fluxes to compute the following line ratios: \oiii/\hb, \nii/\ha, \sii/\ha. We will also show line ratio maps for those 3 line ratios and additionally, for \oi/\ha.

\end{enumerate}

With the LZIFU outputs, we construct maps for all the physical properties listed above (Section~\ref{sec:result}). Each MUSE spaxel spans $0\farcs2$ on a side. We smooth the maps with a 3-spaxel Gaussian kernel, which corresponds to a full-width-half-max (FWHM) $\sim1\farcs4$ and is comparable to the 1\arcsec seeing. We thus maintain a high spatial resolution close to the seeing, thanks to the data having a sufficient signal-to-noise ratio to detect most of the quantities of interest over the majority of the MUSE field-of-view.

\subsection{Velocity and Velocity Dispersion Profiles}\label{sec:profiles}

We extract velocity and velocity dispersion profiles along a few position angles (PAs) including the galaxy PA of 157 degrees East of North, and define two lines along the cone edges with PA=15 and 115. Profiles are computed by running an average from the unsmoothed velocity and velocity dispersion maps and using an average window that achieves the best compromise between high spatial resolution and high signal-to-noise. We employ a 7-spaxel binning size along the PA, and for each bin we compute the average and standard deviation using a width of 7~spaxels perpendicular to the PA. Changing these values slightly does not affect the results significantly. In particular, smaller width and bin size result in noisier measurements (larger standard deviation) but does not alter the normalization and overall shapes of the velocity and velocity dispersion profiles. However, using significantly larger bins artificially flattens the KDC velocities, as one would expect. The results are described in Section~\ref{sec:kin}.

\section{Results}\label{sec:result}

\subsection{Emission-Line Intensity Maps}\label{sec:linemap}

\begin{figure*}
\epsscale{1.1} \plotone{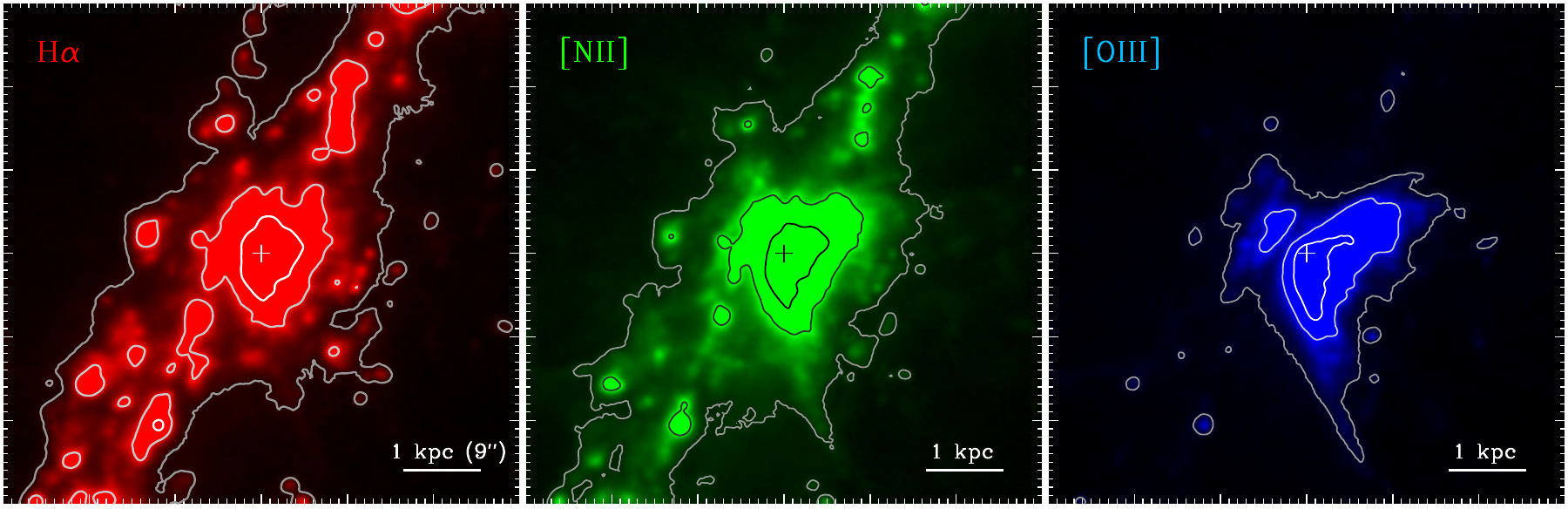}
\caption{Emission-line intensity map for three strong lines as follows: Red = \ha, Green = \nii, Blue = \oiii. For each emission-line, we show the total line flux across the central $60\arcsec \times 60\arcsec$ of the MUSE field-of-view. North is up; East is left. The scale bar measures 9\arcsec, which corresponds to 1~kpc at the distance to NGC~7582. The plus symbol marks the center of the galaxy. All panels have the same color normalization, and each emission-line flux map is encoded in its respective channel of a RGB image (\ha\ in R, \nii\ in G, \oiii\ in B). The contours are spaced logarithmically, with the outermost contour fixed at the same threshold ($5\times10^{-14}$~erg~s$^{-1}$~cm$^{-2}$).
}\label{fig:lineMaps}
\end{figure*}

We produced maps of integrated emission line fluxes from the fits obtained with LZIFU, described in Section~\ref{sec:fit}. We show the spatial distribution of the emission in the following strong lines: \ha, \nii, and \oiii\ (Figure~\ref{fig:lineMaps}). The \ha\ and \nii\ maps show a combination of emission along the galaxy disk (diagonal; with enhanced regions likely corresponding to denser star-forming regions) and from the nucleus and associated ionized cones. In contrast, the \oiii\ map mostly shows emission arising from the ionized cones with very little from the galaxy disk itself. The lack of obvious \oiii\ emission in the star-forming disk can be expected if the gas is significantly metal enriched \citep[e.g.,][]{kew01}, which is indeed what we find given the elevated \nii/\ha\ emission line ratios when constructing emission line ratio maps and diagnostic diagrams (Section~\ref{sec:ratio}).

\begin{figure}
\epsscale{0.85} \plotone{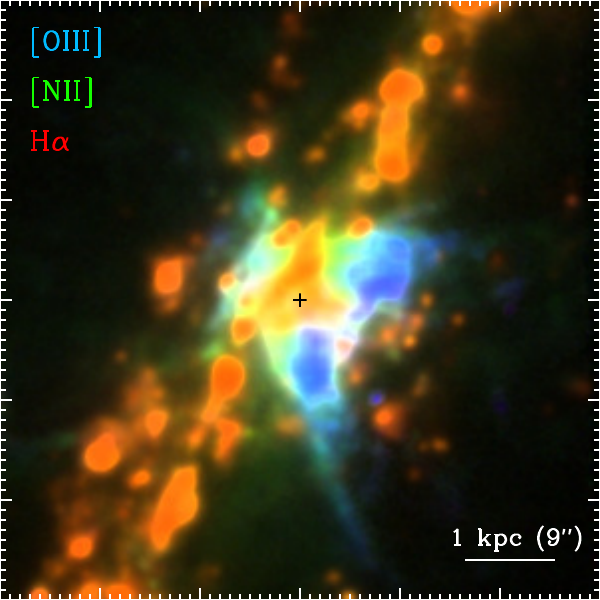}
\caption{ Emission-line intensity color map combining the three strong lines shown in Figure~\ref{fig:lineMaps} as RGB channels (Red = \ha, Green = \nii, Blue = \oiii, shown with a linear scaling of, respectively, 0.9, 1.05, 1.05). North is up; East is left. The plus symbol marks the center of the galaxy. The panel spans the central $60\arcsec  \times 60\arcsec$ of the MUSE field-of-view, with a major tick mark every 10\arcsec.
}\label{fig:lines}
\end{figure}

To compare the spatial locus of the emission from the three spectral lines, we encoded them in the Red, Green, Blue (RGB) channels to create the color image shown in Figure~\ref{fig:lines}. This visualization showcases the relative intensity of the emission, and indicates that \ha\ dominates in the galactic disk, \oiii\ dominates in the ionization cones, and that the central region at the base of the cones might have a mixed contribution including starbursting activity. We will revisit this possibility, which is in agreement with previous work, when we examine the line ratio maps (Section~\ref{sec:ratio}). First, we investigate the dynamical properties of the stellar and gaseous components in order to fold them into the full picture.

\subsection{Stellar and Nebular Gas Kinematics}\label{sec:kin}

\begin{figure*}
\epsscale{1.2} \plotone{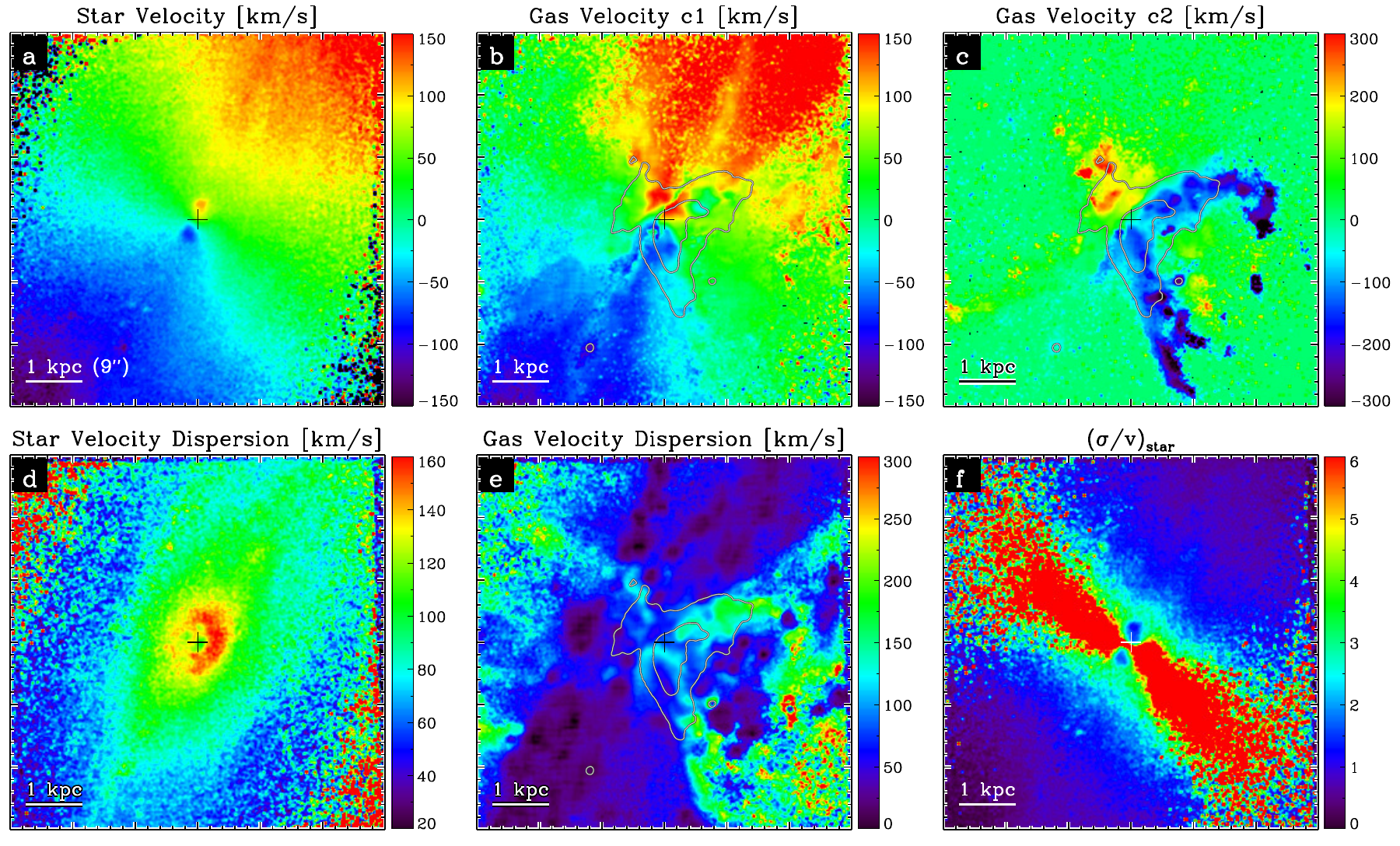}
\caption{
   (a) Stellar velocity map (in \kms) showing a rotating disk on large scales, and a kinematically distinct core with a differential velocity in the same direction as the main disk. (b) Gas velocity map for the first component (in \kms), which is defined as closer to the stellar component when two components were present {for \oiii\ or \ha}. We overlay \oiii\ flux contours for reference. (c) Gas velocity map for the second component (in \kms), which was used when \oiii\ or \ha\ had a significantly detected (S/N$>$3) second component. This kinematic component is largely associated with outflowing cones, which are also traced by the overlaid \oiii\ flux contours. The right-hand side cone is at the front, and flowing toward us while the left-hand side cone is behind the galaxy disk, therefore more strongly attenuated and flowing away from us. (d) Stellar velocity dispersion (in \kms), which rises toward the center, except for a sigma-drop associated with the KDC. (e) Nebular gas velocity dispersion (in \kms). This map has a lot more structure than the stellar velocity dispersion, namely the highest values are reached along the edges of the cones and their large-scale extension. (f) Ratio $(\sigma/v)_{star}$ further highlighting kinematically cold (rotation-dominated) signatures at the KDC location. North is up; East is left. The scale bar measures 9\arcsec, which corresponds to 1~kpc at the distance of the target. Contours show the total \oiii\ flux (smoothed over $3\times3$ spaxels; in log space). The plus symbol marks the center of the galaxy. The panel spans the central $60\arcsec  \times 60\arcsec$ of the MUSE field-of-view, with a major (minor) tick mark every 10\arcsec (2\arcsec).
}\label{fig:vel}
\end{figure*}

The stellar kinematics of NGC~7582 exhibits a regular rotation pattern on the largest scales probed by MUSE ($\sim$8~kpc), spanning rotation velocities from -150 to +150~\kms (Figure~\ref{fig:vel}a). Interestingly, we find a kinematically distinct core (KDC) in NGC~7582, which is co-rotating with the large-scale galactic disk but with faster relative velocities and limited to a diameter around $600-700$~pc.
The presence of a KDC is supported by the sigma-drop signature in the velocity dispersion map (Figure~\ref{fig:vel}d), which shows a mild dip in the stellar velocity dispersion at the location of the KDC instead of a central peak. Beyond the central region, the velocity dispersion slowly decreases with radius, but with some asymmetry in the sense that we find higher values toward the North-West compared to the South-East. The velocity dispersion to velocity ratio for the stellar component (panel f) indicates rotation dominated kinematics, including two kinematically cold spots at the location of the KDC, where the enhanced velocity and mildly suppressed velocity dispersion produce a striking contrast relative to the surrounding regions. The KDC could be either a disk or a ring of stars and gas co-rotating with the main disk but with a differential, faster velocity. We will investigate more closely its kinematics and interpretation with Figures~\ref{fig:velpro} \& \ref{fig:velproZoom}.

\begin{figure*}
\epsscale{1.2} \plotone{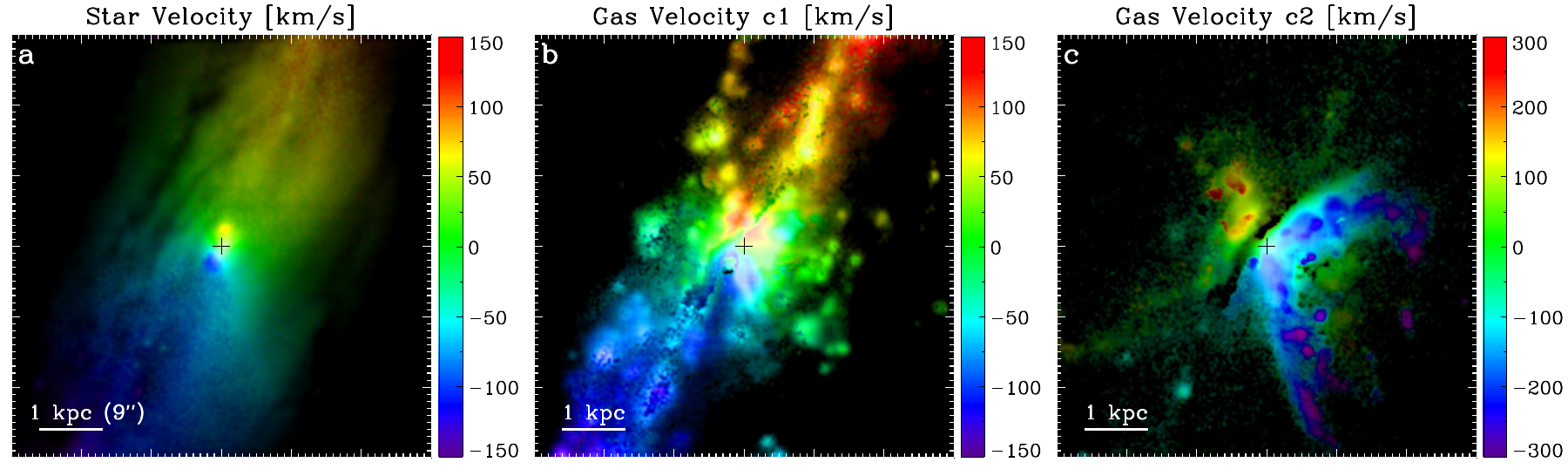}
\caption{
   Light-weighted versions of the velocity maps from Figure~\ref{fig:vel}: (a) Stellar velocity map (in \kms) weighted by the integrated stellar continuum between $5030-8200$~\AA\ (after masking emission lines). (b) Gas velocity map (in \kms) weighted by the \ha\ flux in gas component 1, which is the closest to stellar kinematics. (c) Gas velocity map for component 2, (in \kms), weighted by the \oiii5007\ flux in the second component. North is up; East is left. The scale bar measures 9\arcsec, which corresponds to 1~kpc at the distance of the target. The plus symbol marks the center of the galaxy. The panel spans the central $60\arcsec  \times 60\arcsec$ of the MUSE field-of-view, with a major tick mark every 10\arcsec.
   }\label{fig:velweight}
\end{figure*}

\begin{figure*}
\epsscale{1.2} \plotone{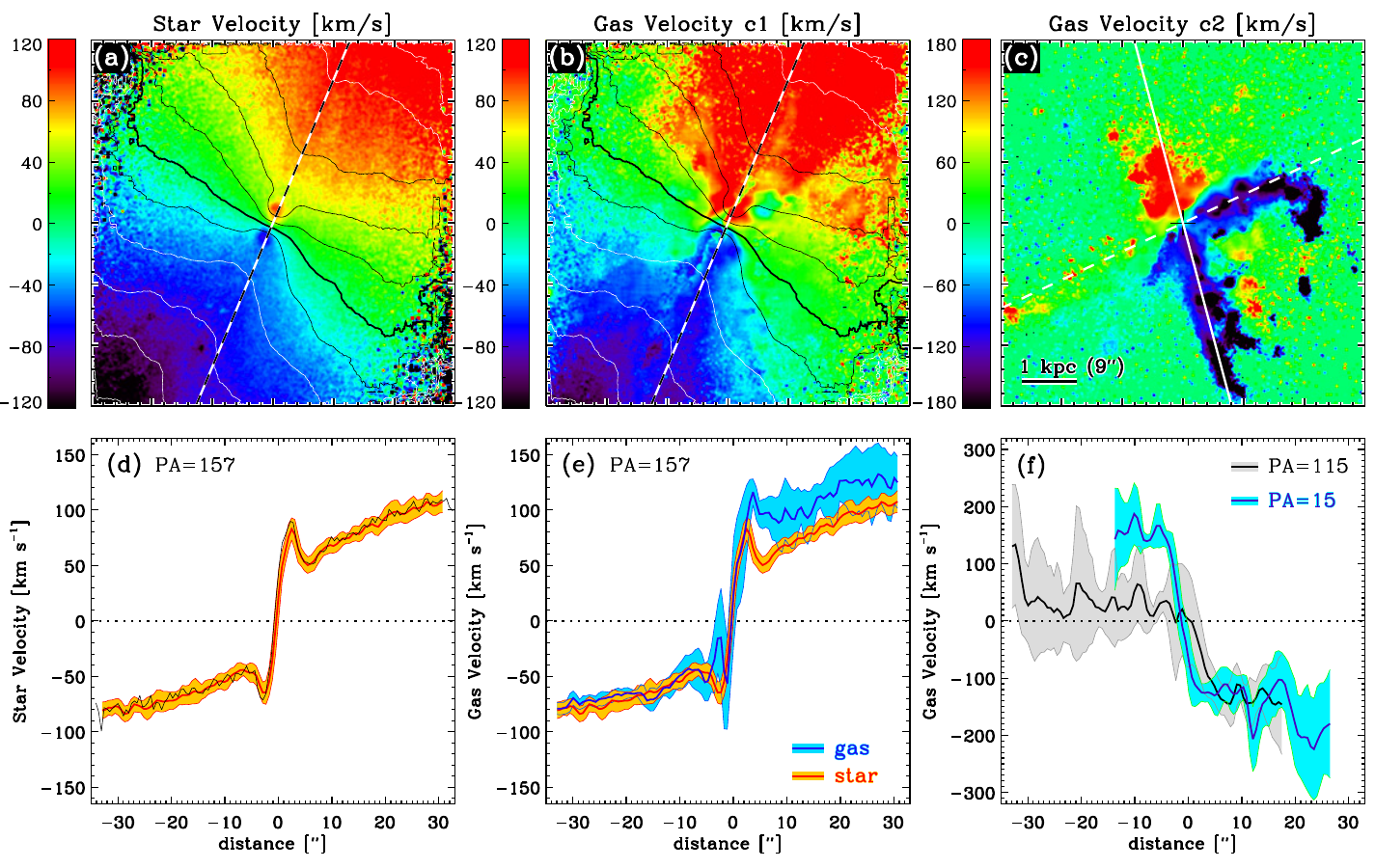}
\caption{
   Top row: velocity maps similar to the top row of Figure~\ref{fig:vel} except that here we show iso-contours of the stellar velocity field, the PA of 157 (dashed line), and an enhanced velocity contrast.
   (a) Stellar velocity map (in \kms); (b) Gas velocity map (in \kms) for Component 1; (c) Gas velocity map (in \kms) for Component 2. North is up; East is left. The scale bar measures 9\arcsec, which corresponds to 1~kpc at the distance of the target.
   Bottom row: velocity profiles (in \kms) along the PA=157 for the stellar and gas components. 
   (d) Stellar velocity profile, which shows a clear velocity excess at the KDC while the main curve is slowly rising in amplitude with distance away from the center. The thick red line and shaded region correspond to the running average and standard deviation, while the thin black line is a simple cut along the PA of the Gaussian-smoothed map shown to highlight the peaks in the KDC velocities. (e) Nebular gas velocity profile for Component 1, where the thick blue line and shaded region show the running average and standard deviation, compared to the stellar profile from panel d (thick red line with orange shaded region).  (f) Nebular gas velocity profile for Component 2 along axes defined to follow the edges of the cones with PA=15 (purple line with cyan shaded region; solid line in panel c), and PA=115 (black line with grey shaded region; dashed line in panel c).
}\label{fig:velpro}
\end{figure*}

\begin{figure*}
\epsscale{1.2} \plotone{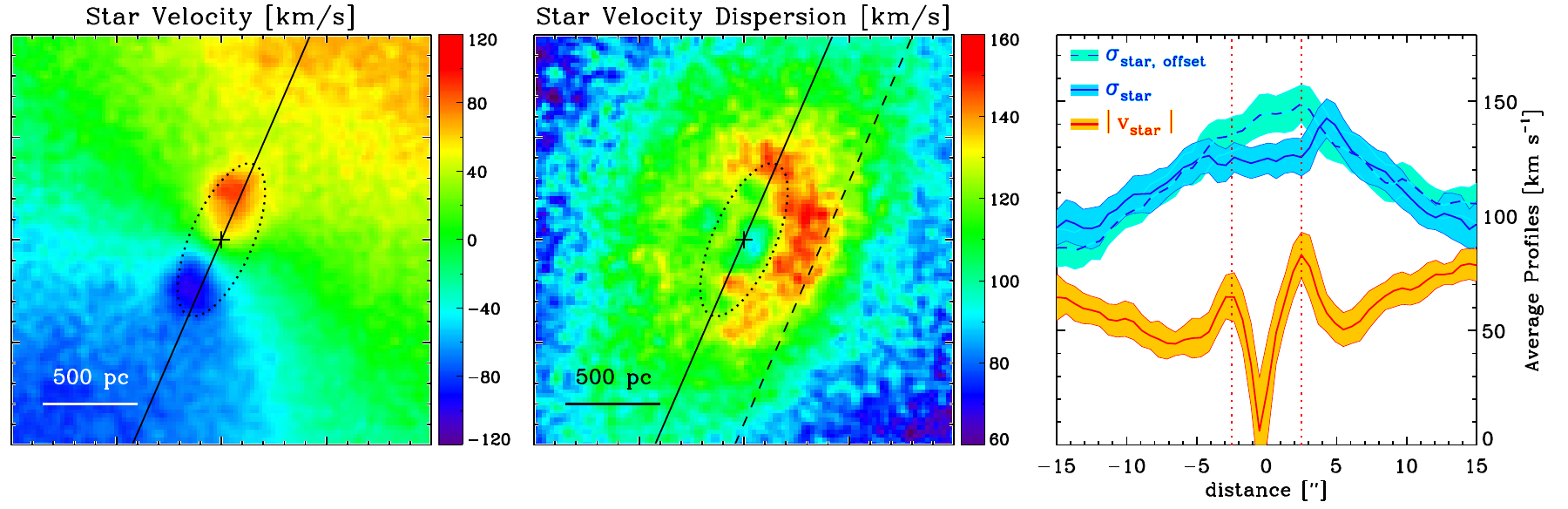}
\caption{
 Zooming on the central $20\arcsec \times 20\arcsec$ of the MUSE field-of-view to compare the stellar velocity map (left panel), stellar velocity dispersion map (middle panel), and their average profiles along the galaxy PA of 157 degrees. The PA is drawn on the first two panels (solid black line) as well as an ellipse illustrating the extent of the KDC with major and minor axes of 8\arcsec and 3\arcsec, respectively (dotted line). We also computed the velocity dispersion profile offset by 3\farcs5 at the location shown by the dashed line (middle panel). This zoomed view clearly shows the sigma-drop at the location of the KDC. We show the absolute value of the velocity profile and standard deviation (red line with orange shaded area), and highlight the two peaks of the KDC velocities with vertical dotted lines at $\pm 2\farcs5$ from the center. The stellar velocity dispersion is shown in blue and exhibits a lower value at the location of the KDC than would be expected for a dispersion-dominated profile that peaks toward the center. As a comparison, the velocity dispersion profile offset by 3\farcs5 (blue dashed line with cyan shaded region) more centrally peaked.
}\label{fig:velproZoom}
\end{figure*}

The gas kinematics was decomposed into two components modeled with two Gaussian profiles, fit to the emission lines as described in Section~\ref{sec:fit}. We recall that the first component (c1) was assigned to the closest velocity to the stellar kinematics. We recover a similar large-scale, regular rotation field for the c1 gas velocity component (Figure~\ref{fig:vel}b) as for the stellar velocity map with hints of additional structure and possible deviations from a pure rotating disk. The c2 kinematic component instead reveals outflowing cones with a blue-shifted front cone, and a red-shifted back cone (Figure~\ref{fig:vel}c)\footnote{The two-component fits result in faint emission (but with S/N$>3$ in \ha\ and/or \oiii) outside the cones with a range of velocities, including near-zero velocities. This comparatively very faint signal is sufficient to fill in the velocity map shown in Figure~\ref{fig:vel}(c) after smoothing. However, it is not significant when considering the flux-weighted version of the same map (Figure~\ref{fig:velweight}c), so we do not consider it as physically significant in our interpretation and analysis of the gas kinematics.}. The cones exhibit a stronger signal along the edges, which we attribute to limb brightening, and which suggests that the central portion of the cone is more diffuse than the edges, or that the cone may be nearly hollow. The redshifted back cone appears faint in \oiii\ emission (contours on panels b \& c) due to heavy dust attenuation from the host galaxy. This is not surprising given the obvious dust lanes at that location, which we investigate in the next section (Figure~\ref{fig:dust}a). 

In order to combine information from the kinematics and from the signal strength, we generated light-weighted versions of the velocity maps (Figure~\ref{fig:velweight}). Respectively, we weighted the stellar velocity map with the integrated continuum light between $5030-8200$~\AA (panel a), we weighted the first gas velocity component map, primarily tracing the galaxy disk, with the integrated \ha\ line flux (panel b), and we weighted the second gas component velocity map with the integrated \oiii\ line flux (panel c). The weighting works by coding each spaxel from black to white (value) based on the signal strength while the hue is still attributed to the velocity values from purple (approaching) to red (receding). We notice on the stellar velocity map that the KDC is bright relative to the surrounding disk. The first gas component (c1) map shows a somewhat clumpy structure with knots that likely correspond to regions or knots with enhanced star formation activity.  There is also potentially some differential or shear velocity structure along the semi-major axis, which could be due to the (known) presence of a bar in this galaxy \citep{mor85}. The second gas component (c2) map highlights a combination of velocity structure, and of emission line flux variations along and within the outflowing ionized cones. The upper ridge of the front cone shows slightly lower projected velocity, which could tentatively correspond to the gas slowing down as the cone expands and hits surrounding medium. In this case, we might expect signatures of shocked gas along and outside the outer ridges of the cones.

The stellar and gas kinematics are displayed with enhanced contrast and with iso-contours in the top row of Figure~\ref{fig:velpro}, where we also draw lines along the major axis of NGC~7582 reported to have a position angle (PA) of 157 \citep{jar03} in panels (a) and (b). Velocity profiles are computed from a running average and standard deviation along the PA as described in Section~\ref{sec:profiles}. The resulting profiles are displayed in the bottom row, where the filled area encompasses the standard deviation around the average (solid line). In the case of the stellar velocity map, we also show a velocity profile along a single line, which we obtain by reading the value from the nearest spaxels that are crossed by the line on the smoothed map shown in panel (a). This single-line profile (thin black line in Figure~\ref{fig:velpro}d) is shown for comparison, and demonstrates that the KDC profile is narrowly peaked while being contained within the standard deviation (orange filled area) of the stellar velocity profile obtained with the running average (thick red line). 

The stellar and gas velocity profiles support the presence of the KDC, with a steep gradient from -70 to +90~\kms\ over $<5$\arcsec ($\sim 550$ pc). Relative to the stellar velocity profile, extracting velocity profiles along the gas kinematic maps result in larger variations (standard deviations). For component c1, we show a direct comparison between the velocity profiles of the gas (in blue) and the stars (in red) in panel (e). The average trends show a systematic difference on the NW (redshifted) side in the sense that the gas is rotating faster than the stars by $\sim40$\,\kms, while both velocity profiles are consistent with each other within the uncertainties on the SE (blueshifted) side.

For the second component of the gas (c2), we probe the outflowing cones, and obtain velocity profiles along both edges at PA=15 and 115 (lines shown in panel c). The average and standard deviation were computed over a width of $3\farcs6$ perpendicular to the PA to account for spatial variations across the cone edges, and 7 spaxels (1\farcs4) along the PA to maintain the same sampling along the profiles as in panels (d) and (e). The cone profiles are characterized by very sharp velocity jumps around the central position. The back cone is significantly detected along the PA=15 angle, and shows similar projected velocity amplitude as the front cone reaching 150-200~\kms. This is consistent with a symmetrical biconical outflow. Along PA=115, the outflow is detected significantly for the front cone, but not for the back cone. On the c2 gas velocity maps, there are some small clump-like regions with redshifted velocities near the dashed line (PA=115) but the computed average velocity profile is noisy, and only displays a small systematic enhancement on the redshifted side in panel (f). Given that the back cone is behind the galaxy disk along our line-of-sight, it is possible that we miss some of the c2 gas component of the back cone due to obscuration, which is slightly higher in the South-East quadrant of the field-of-view (Section~\ref{fig:dust}). In panel (c), we also see faint but significant redshifted regions within the edges of the front cone. We hypothesize that these regions may correspond to the back side of the front cone, which could be seen if these regions are denser and/or more luminous than average and assuming a cone with a hollow or very diffuse inner core. Overall, we distinguish three kinematic components: (i) a galactic disk with regular rotation of both stars and gas; (ii) a co-rotating KDC with diameter $\sim$600pc; and (iii) a biconical gas outflow extending over $>$3~kpc. 

In Figure~\ref{fig:velproZoom}, we zoom in the central $20\arcsec \times 20\arcsec$ region to examine more closely the KDC velocity profile, as well as the velocity dispersion profiles over the KDC along the PA (solid line) compared to a velocity dispersion profile along the same angle but offset from the KDC (dashed line). In this closer view, we draw an ellipse for visual reference to help compare the velocity map (left-hand panel) and velocity dispersion map (middle panel). The average velocity and velocity dispersion profiles are computed with a running mean and standard deviations over $7 \times 7$ spaxels perpendicular and along the PA (solid line), and shown in the right-hand panel.  The main results from these three panels is that the velocity signature of the KDC is accompanied by a suppressed velocity dispersion (sigma-drop) relative to a dispersion dominated system that would peak at the center. By comparison, the velocity dispersion profile obtained slightly offset (by 3\farcs5) from the KDC shows a more centrally peaked shape, and the difference suggests that the sigma-drop is around $20-30$\,\kms.

Those signatures clearly indicate a dynamically cold, rotation-dominated structure for the KDC, such as a ring or disk. The ring interpretation is consistent with high-resolution Atacama Large Millimeter/submillimeter Array (ALMA) observations of the CO(3-2) transition, which revealed a molecular gas ring with a diameter $\sim 400-600$~pc \citep[][also see discussion in Section~\ref{sec:kdc}]{alo20,gar21}. But it is also possible that the KDC is a nuclear stellar disk that is bounded by the ring of molecular gas, as may occur in strongly barred spiral galaxies \citep{gad20}, in which case the nuclear stellar disk is characterized by younger stellar ages relative to the main galaxy \citep{bit20}. The MUSE observations do not resolve an inner opening, but we will keep in mind the presence of a molecular gas ring component for the KDC as part of the physical interpretation.

\subsection{Dust Attenuation}\label{sec:dust}

\begin{figure*}
\epsscale{1.} \plotone{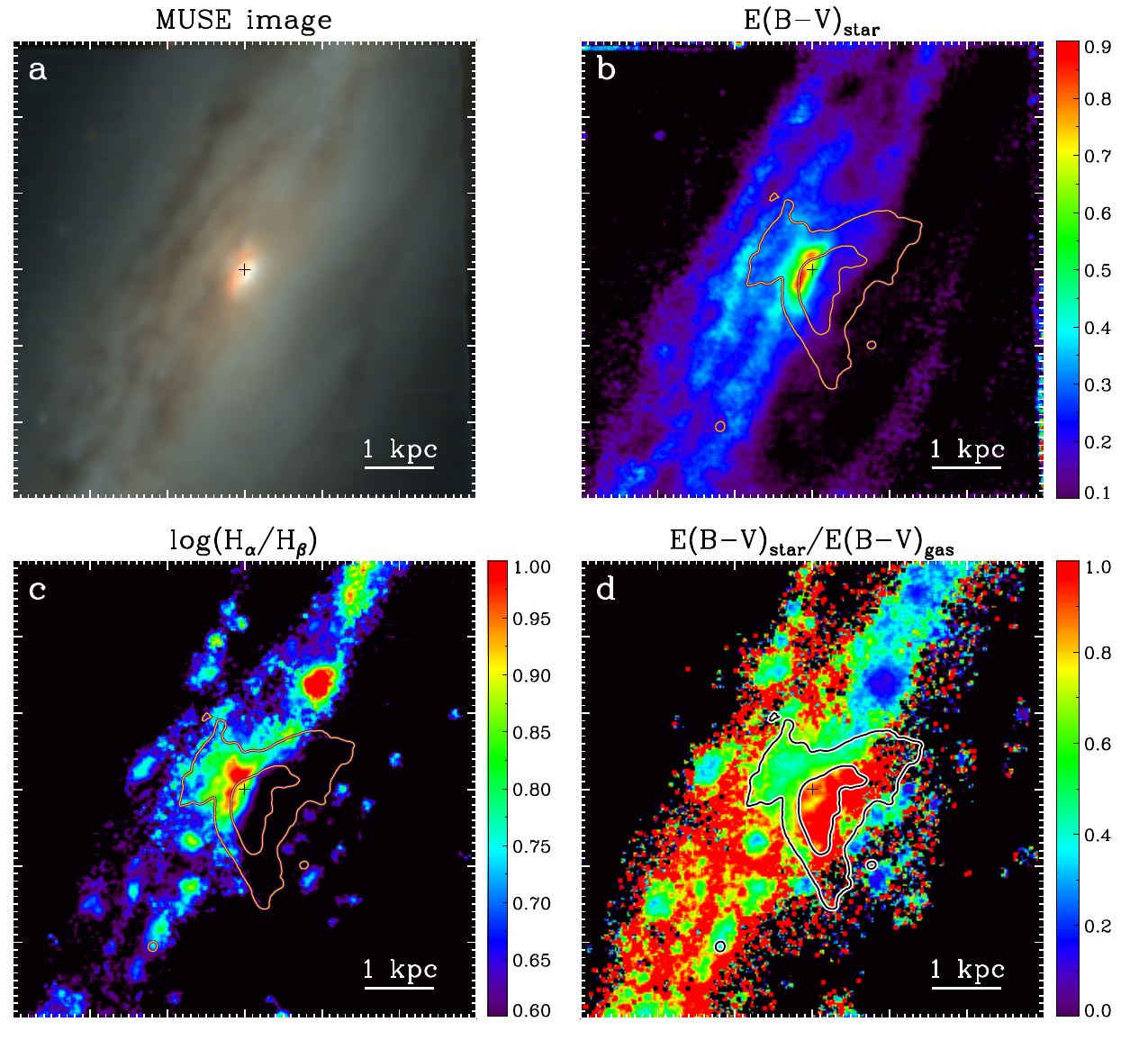}
\caption{
   (a) RGB color image of NGC~7582 from the MUSE stellar continuum using three synthetic bands defined to avoid emission lines and sky line residuals (central wavelengths are: $B\sim$5400~\AA, $G\sim$6100~\AA, $R\sim$7400~\AA). 
(b) Stellar reddening, E(B-V), inferred from the stellar continuum modeling. The main feature is an elongated peak of extinction in the central region. (c) Gas attenuation traced by the Balmer Decrement \ha/\hb. A larger value indicates a more important dust obscuration. The appearance is more patchy than that of the stellar reddening from panel (b). In addition to the strong obscuration in the central region, there is another peak toward the North-West with no obvious counterpart in the stellar absorption. (d) Ratio of star-to-gas E(B-V). For reference, the Calzetti value is 0.44 (green). We find spatial variations of this ratio toward both higher star-to-gas ratio (e.g., in diffuse disk regions, and a portion of the front cone appearing red), and lower ratios (e.g., in particular toward the North-West in two regions colored in blue). North is up; East is left. The scale bar measures 9\arcsec, which corresponds to 1~kpc at the distance of the target. Contours show the total \oiii\ flux (smoothed over $3\times3$ spaxels; in log space). The plus symbol marks the center of the galaxy.
}\label{fig:dust}
\end{figure*}

From the MUSE data cube, we generated three synthetic bandpasses in order to produce a RGB color map. We defined three spectral ranges within which only pixels reliably tracing the stellar continuum were used, e.g., avoiding strong sky lines and emission lines. The central wavelengths are effectively: $B\sim$5400~\AA, $G\sim$6100~\AA, $R\sim$7400~\AA. Their bandpass widths were defined to ensure that each band includes a similar number of valid spectral pixels ($\sim$450). The individual color channels were combined with an asinh scaling. The resulting RGB color map (Figures~\ref{fig:labels} and \ref{fig:dust}a) shows the bright nucleus, and several dust lanes over the stellar continuum. The region corresponding to the 600~pc KDC appears heavily reddened on this map, indicating an important foreground dust screen in front of a light source.

We generated dust attenuation maps from both the stellar and gas components. From the former, we use the best-fit reddening, E(B-V), from the LZIFU fit to the stellar continuum (Figure~\ref{fig:dust}b). The stellar reddening indicates the presence of dust lanes along the disk, and is characterized by a strong peak near the base of the \oiii\ front cone (contours). From the latter, we construct a map of the \ha/\hb\ Balmer Decrement (Figure~\ref{fig:dust}c). Dust-free regions have values of 0.45-0.5~dex (\ha/\hb=2.86-3.1) while deviations toward higher ratios indicate higher dust attenuation. The Balmer Decrement map also exhibits a peak of dust attenuation along the base of the front cone. It otherwise has a more clumpy appearance than the stellar attenuation map, and, intriguingly, shows two regions with high gas extinction North-West of the center that are not seen in the stellar attenuation map.

To compare the stellar reddening to the gas extinction, we converted the Balmer Decrement to E(B-V)$_{gas}$ assuming an intrinsic ratio of 2.86 \citep[$n_e=100~$cm$^{-3}$; $T_e=10^4$~K]{ost89}, and the Calzetti extinction law \citep{cal00}. The resulting stellar-to-gas reddening ratio map (Figure~\ref{fig:dust}) shows dense regions with higher gas-to-stellar extinction surrounded by diffuse medium with lower gas-to-stellar extinction. Several regions are fairly close to the canonical Calzetti value of 0.44 (green color). The Northern side has elevated gaseous attenuation relative to stellar attenuation, which is particularly noticeable around two high attenuation blobs noted above with high Balmer Decrements. The underlying cause for these features is not fully clear though they plausibly correspond to regions of dusty star-formation so obscured that the background continuum signal is completely missing (optically thick), and therefore not accounted for in the spectral fitting used to determine the stellar reddening. They could also correspond to young \hii\ regions that lack substantial stellar mass content and are dominated by the light of young dust-embedded stars still surrounded by their birth cloud.

Overall, the common features of both the stellar and gas attenuation maps are the peak attenuation in the central region corresponding to the KDC and to the base of the front \oiii\ cone, as well as the large scale, somewhat clumpy, dust lanes. The back cone suffers from heavier attenuation compared to the front cone. For the former, it appears that the emission line signal is dominated by low-attenuation gas that shines between the large-scale dust lanes, as shown in the bottom center and bottom left panels of Figure~\ref{fig:labels}, where we added the \oiii\ emission in the green channel of the reconstructed RGB image.

The peak obscuration on the KDC corresponds to values of stellar reddening reaching up to $E(B-V)_{star}=0.86$. From their sample of lightly reddened quasars, \citet{wil04} found that quasars with $E(B-V)\sim0.5$ span a range of $N_H$ from $3\times10^{21}-4\times10^{23}$~cm$^{-2}$, which place them closer to a SMC-like gas-to-dust ratio \citep[$N_H = 5.2 \times 10^{22}\times E(B-V)$;][]{bou85} than to a MW-like ratio. Assuming a SMC-like gas-to-dust ratio, the peak extinction corresponds to $N_H\sim1.3\times10^{23}$~cm$^{-2}$. 

From the Balmer Decrement, the nebular gas attenuation reaches $E(B-V)_{gas}=0.97$ over the KDC, assuming the Calzetti attenuation law as described above. Assuming a SMC-like gas-to-dust ratio yields $N_H\sim2.1\times10^{23}$~cm$^{-2}$. This value is comparable to the estimate from the stellar reddening above. Using a MW-like gas-to-dust ratio would give lower gas column densities by nearly a factor of ten. On small spatial scales, the above should be regarded as a lower limit, as smaller clumps below the resolution limit reach much higher column densities, blocking completely a fraction of the starlight and/or nebular emission \citep[e.g.,][]{marko14}. This is particularly relevant toward the AGN, for which the cross-section is unresolved, and which is located within the KDC. While there remains uncertainty due to the choice of gas-to-dust conversion factor, our estimate is likely conservative in the sense that the true column density along the line of sight to the nucleus could be even higher. 

\subsection{Gas Excitation}\label{sec:ratio}

\subsubsection{Emission-Line Ratio Maps}

We detected emission from ionized gas in several spectral lines. A composite map of three strong lines (\ha, \nii, and \oiii; Figure~\ref{fig:lines}) gives an overview of the dominant sources of ionization. The main disk of NGC~7582 harbors a large number of star-forming knots or complexes, which are mostly emitting in \ha. The central region is the most mixed with strong emission in all lines, while the cones are dominated by emission in \oiii, and, to a lesser degree, \nii.

\begin{figure*}
\epsscale{0.8} \plotone{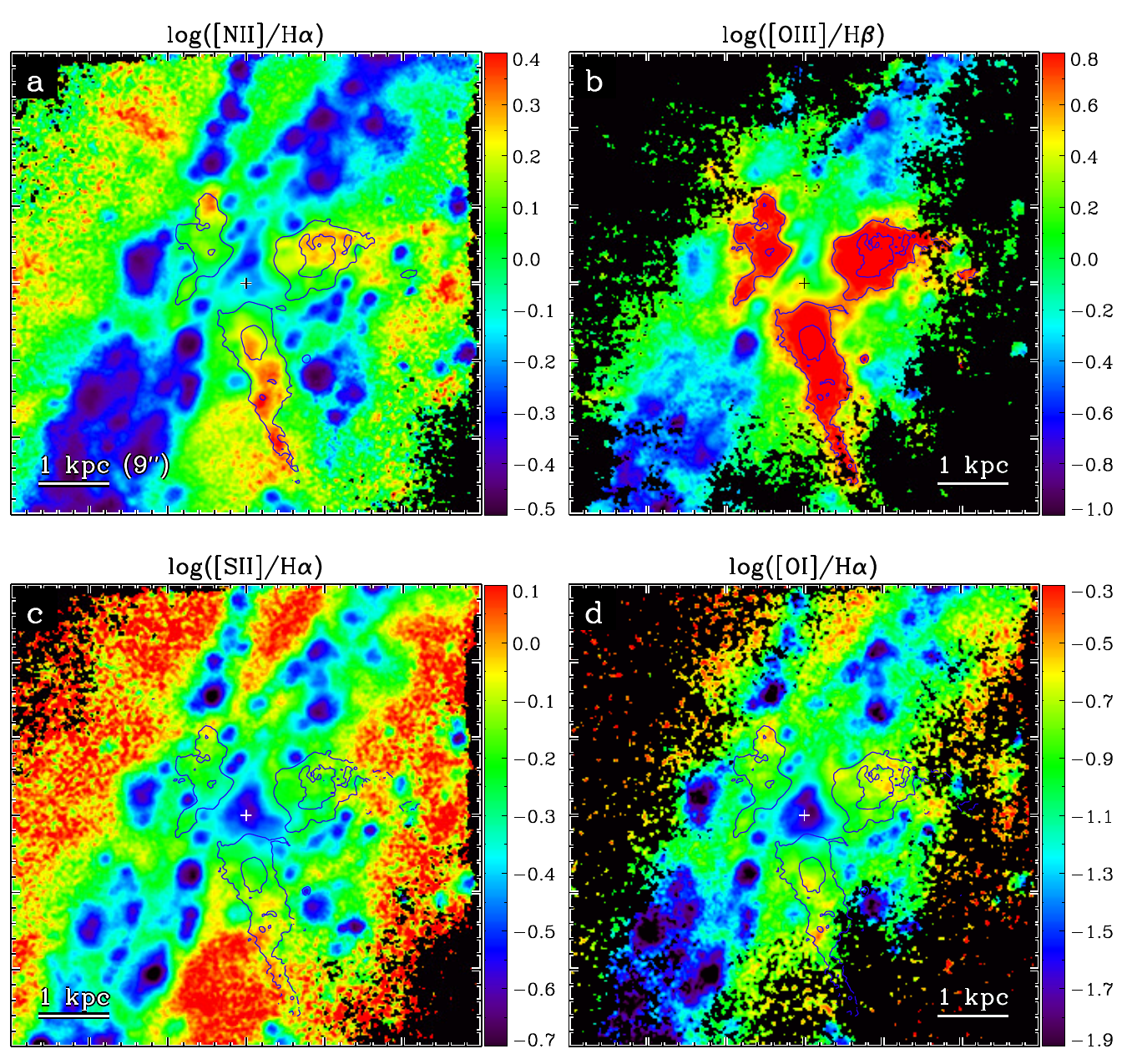}
\caption{ Emission-line ratio maps. (a) \nii/\ha\ (b) \oiii/\hb\ (c) \sii/\ha\ and (d) \oi/\ha. The values of the logarithm of the line flux ratios is color-coded as indicated with each color bar. The overlaid contours correspond to the region with the highest \oiii/\hb\ values from panel (b). North is up, East is left. The scale bar measures 9\arcsec, which corresponds to 1~kpc at the distance of the target.
}\label{fig:ratio}
\end{figure*}

\begin{figure*}
\epsscale{0.8} \plotone{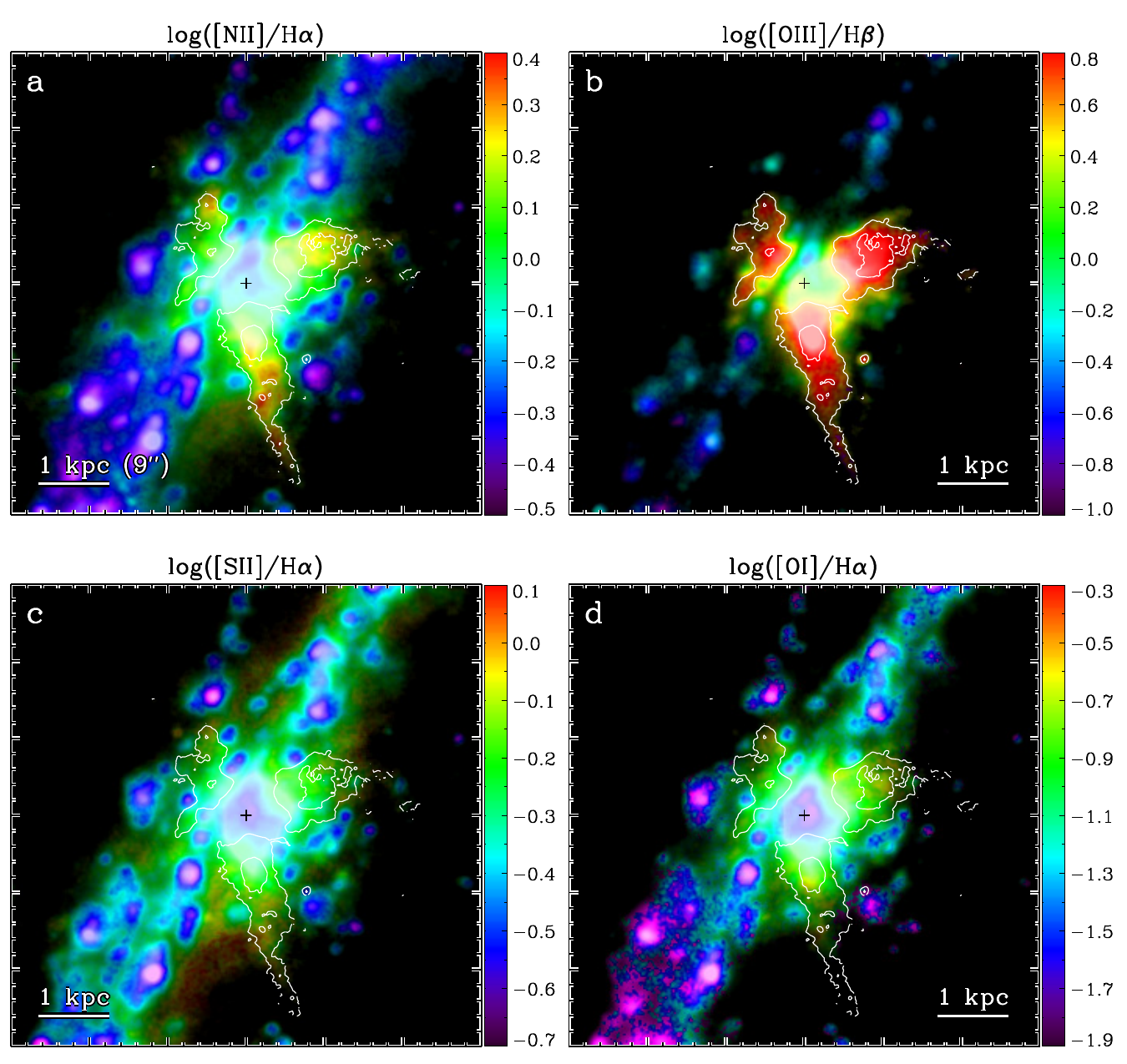}
\caption{ Same emission-line ratio maps as in Figure~\ref{fig:ratio}, except that each spaxel is now flux weighted using the sum of line fluxes for each ratio. (a) \nii/\ha\ (b) \oiii/\hb\ (c) \sii/\ha\ and (d) \oi/\ha. The values of the logarithm of the line flux ratios is color-coded as indicated with each color bar. The overlaid contours correspond to the region with the highest \oiii/\hb\ values from panel (b). North is up, East is left. The scale bar measures 9\arcsec, which corresponds to 1~kpc at the distance of the target.
}\label{fig:ratio2}
\end{figure*}

To gain additional insight about gas excitation conditions, we constructed line ratio maps of \nii/\ha, \oiii/\hb, \sii/\ha, and \oi/\ha\ (Figures~\ref{fig:ratio} \& \ref{fig:ratio2}). 
Regions dominated by star formation exhibit \hii-region like ratios with low values of \nii/\ha, \sii/\ha, and \oi/\ha\ (purple to blue colors on the line ratio maps). 
We can see a number of star-forming region complexes distributed along the main galaxy disk with a clumpy appearance. Those are surrounded by regions with 
slightly more elevated line ratios (green color on the maps), perhaps dominated by diffuse WHIM (warm hot interstellar medium). On the \oiii/\hb\ map, there are two very striking, highly ionized cones. In contrast, the clumpy, star-forming regions have very low values of \oiii/\hb, likely indicating a combined effect of comparatively higher metallicity, and lower ionization parameter citep{kew19}.

We highlight the regions with the most elevated \oiii/\hb\ ratios with contours. They correspond to the locus of the front and back \oiii\ outflowing cones. The back cone stands out more strongly here than with the \oiii\ contours alone. Thanks to the close proximity in wavelength of the \oiii\ and \hb\ emission lines, they will suffer from the same dust obscuration if they arise from the same physical regions. Therefore, even if both lines are strongly attenuated, their ratio keeps the physical signature of the source of ionization. This result strengthens the interpretation of the presence of the back cone, and its association with a biconical ionized gas outflow. 

To convey the relative importance of the ionized gas signal with the associated line ratios, we introduce a line flux weighting scheme. For each spaxel of Figure~\ref{fig:ratio2}, the hue corresponds to the line ratio as in Figure~\ref{fig:ratio} and as indicated by the color bars. The lightness is coded with the sum of the line fluxes for a given line ratio. This view of the line ratio maps shows that the bulk of the \oiii/\hb\ signal comes from the cones, while it is shared between the star-forming knots in the galactic disk and the cones for the three other line ratios. While Figure~\ref{fig:ratio} included extended regions with elevated \sii/\ha\ perpendicular to the disk, the line flux is clearly sub-dominant in these diffuse and extended regions.

While we clearly see elevated \oiii/\hb\ ratios along the ionized gas cones, it can be challenging to distinguish between AGN photoionization and fast shocks ($> 500$~\kms) because the combined shock and precursor ionized gas can mimic AGN-like ratios in \oiii/\hb, \nii/\ha, and \sii/\ha\ \citep{all08}. However, the \oi/\ha\ ratio is predicted to differ and, at a given \oiii/\hb, reach higher values for fast shock+precursor compared to AGN photoionization. Slow shocks are more easily distinguishable from Seyfert-like ratios because they tend to exhibit LINER-like ratios with comparatively high \sii/\ha\ and \oi/\ha\ signatures in addition to elevated \nii/\ha\ ratios. If the primary ionization along the cones came from slow shocks, we would expect elevated \sii/\ha\ and \oi/\ha\ along the cones, and if it came from fast shocks + precursor, we would still expect high \oi/\ha\ ratios \citep[][their Figure 33]{all08}. However, neither case is supported by the maps shown in Figure~\ref{fig:ratio}, thus we favor the AGN photoionization scenario.

\subsubsection{Emission-Line Ratio Diagnostic Diagrams}

To further examine the sources of ionization, we construct the BPT \citep{bpt} and VO87 \citep{vei87} emission line diagnostic diagrams, which consist in \oiii/\hb\ as a function of, respectively, \nii/\ha\ and \sii/\ha. We will interpret them together with the gas velocity dispersion measurements, and location in the galaxy to further distinguish between AGN photoionization and fast shocks. This is motivated by the work of \citet{dag19}, who reported that shock-ionized regions tend to be characterized by broader velocity dispersions compared to AGN-photoionized regions \citep[also see ][for a review on emission line physical interpretation]{kew19}.

\begin{figure*}
\epsscale{1.15} \plotone{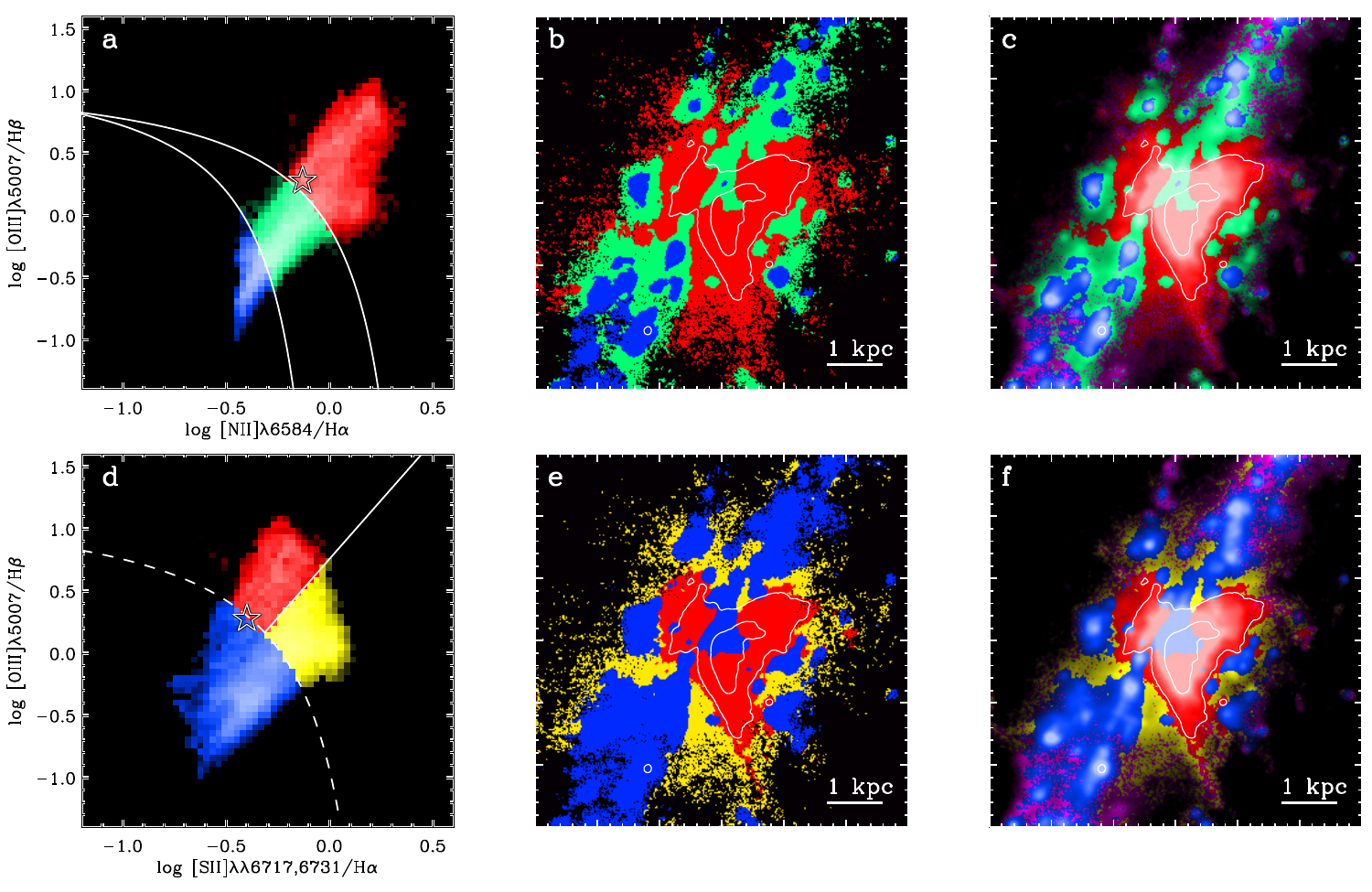}
\caption{ (a) BPT emission line diagnostic diagram. Points are color-coded according to their location with respect to the dividing lines: blue in the star-forming region below the Ka03 line, green in the AGN/SF composite region between the Ka03 and Ke01 lines, red in the AGN-dominated region above the Ke01 line. Each bin on the BPT is flux-weighted by the sum of the 4 line fluxes (from black to white). The star symbol marks the global line ratios from the MUSE cube. (b) Spaxels color-coded according to the location on the BPT diagram from the first panel. AGN-like line ratios are predominantly observed along the cones, while composite-like line ratios are located in the diffuse regions between the star-forming clumps, and extend in the disk over a diameter of $\sim$6-7~kpc. The star-forming clumps along the disk are the only component with SF-like line ratios. (c) Same as panel (b) except with flux-weighted spaxels, where the sum of all four BPT line fluxes are used for the weights. (d) VO87 emission line diagnostic diagram. Points are color-coded according to their location with respect to the dividing lines: blue in the star-forming region, red in the Seyfert branch, and yellow for the LINER branch \citep[dividing lines from][]{kew06}. Each bin is flux-weighted by the sum of the 4 line fluxes (from black to white). The star symbol marks the global line ratios from the MUSE cube. (e) Spaxels color-coded according to the location on the VO87 diagram from panel (d). Most of the Seyfert-like emission is confined in the cones, while the LINER-like emission traces the interface between the cones and the galaxy disk. The bulk of the disk has line ratios in the SF region of the VO87 diagram. (f) Same as panel (e) except with flux-weighted spaxels, where the sum of all VO87 line fluxes are used for the weights.
The overlaid contours correspond to \oiii\ flux. North is up, East is left. The scale bar measures 9\arcsec, which corresponds to 1~kpc at the distance of the target. The maps in panels  cover the central $60\arcsec \times 60 \arcsec$ of the MUSE field-of-view, with a major tick mark every 10\arcsec.
}\label{fig:bpt}
\end{figure*}

The BPT diagram is displayed in Figure~\ref{fig:bpt} as a flux-weighted map of the spaxels (panel a). The star-forming branch below the \citet[][hereafter Ka03]{kau03c} line is only occupied at the metal-rich end with comparatively high values of \nii/\ha. There are two branches in the AGN region reaching above the \citet[][hereafter Ke01]{kew01} line: one with more elevated \oiii/\hb\ values along the expected Seyfert 2 branch (higher and to the left), and a second branch with lower \oiii/\hb\ but higher \nii/\ha, following LINER-like emission. The latter could be due to shocks or other sources of photoionization. We therefore also consider the VO87 diagram, which is more sensitive to the split between Seyfert-like (above the solid line in panel d, in red) and LINER-like emission \citep[below the solid line on panel d, in yellow;][]{kew06}. 

In panels b \& c of Figure~\ref{fig:bpt}, we color-code the spaxels according to the region occupied on the BPT diagram (panel a). Spaxels falling in the star-forming, composite, or AGN regions of the BPT are respectively displayed in blue, green, and red. Panel c is further light-weighted by the total emission line flux in each spaxel from the addition for the four BPT lines. Similarly, panels e \& f show the MUSE field-of-view with spaxels color-coded according the to VO87 classification: star-forming in blue, LINER-like in yellow, and Seyfert-like in red. We again show both an unweighted map (panel e), and a map weighted by total line flux (panel f). Examining the trends on panels c and f, we note that the BPT diagram is more sensitive to composite emission line ratios from the diffuse gas between the star-forming regions; while the VO87 diagram shows a more pronounced contrast between AGN photoionization over the ionized cones (red) compared to LINER-like emission outside and/or around the cones (yellow).

As a comparison, \citet{dav16} used S7 observations to create spatially-resolved BPT and VO87 diagrams (similarly to Figure~\ref{fig:bpt}a,d), and used grids of AGN and \hii-region photoionization models from {\sc Mappings}\footnote{Using version 5.0 of Mappings: https://miocene.anu.edu.au/Mappings/} \citep[Sutherland et al, in prep.,][]{dop13} in order to investigate the source of pressure in the extended narrow line regions (ENLRs) of four AGN hosts including NGC~7582. They found that NGC~7582 displays a classical mixing sequence from AGN-dominated to starburst-dominated ionization but with significant dispersion. The authors tentatively attributed this dispersion to variations in the gas-phase metallicity and ionization parameter within the ionization cone \citep[e.g.,][]{kew13a}. 
The interpretation of a classical mixing sequence favors AGN radiation as the dominant source of static pressure in the ENLRs of NGC~7582 \citep[rather than gas pressure;][]{dop02,gro04a}, supporting a scenario where the AGN can therefore drive galactic-scale outflows.

The mixing sequence that \citet{dav16} reported (their Figure~2) is comparable to the left-hand AGN branch that we find on the MUSE spatially-resolved BPT (Figure~\ref{fig:bpt}a). However, the S7 observations did not reveal the secondary, right-hand AGN branch on the BPT, which comprises spaxels that occupy the LINER region of the VO87 diagram (Figure~\ref{fig:bpt}d). This difference is likely due to two reasons. First, the smaller S7 field-of-view (38$\times$25~arcsec$^2$) compared to the MUSE field-of-view (1~arcmin$^2$) has a more limited coverage of the extended regions where we find the LINER-like emission line ratios in this work (in yellow in Figure~\ref{fig:bpt}e), which are largely outside the ionization cone (\oiii\ contours). Second, the lower spatial resolution ($\sim$2-3\arcsec) of the S7 observations may result in more blending of adjacent regions and dilute the faint LINER-like signatures on the outskirts of the ionization cone. We posit that the interpretation from \citet{dav16} regarding the important role of AGN radiation pressure applies to the ionization cone itself, although the MUSE observations presented here reveal an additional ionization regime outside and beyond the cone. \citet{lop20} also presented a spatially-resolved BPT analysis of NGC~7582 as part of the AMUSING++ nearby galaxy compilation. They compared with the fiducial {\it AGN-ionized} and {\it shock-ionized} bisector proposed by \citet{shar10}, finding that the measurements from the ionized cone mostly lie in the AGN-wind region (their Appendix~D).

Another method to distinguish between AGN photoionization and shock ionization using integral field spectroscopy was proposed by \citet{dag19}, who designed a 3D diagnostic diagram by combining (i) an emission line ratio function (ELRF) \citep[similar to the mixing sequence from, e.g.,][]{dav16}, (ii) the distance of the spaxels from the center (radius), and (iii) the gas velocity dispersion for the components within the spaxel. The authors found that while both AGN and shock ionization can produce high values of ELRF and be located at small radii, they can be separated using the velocity dispersion values due to AGN photoionized components typically having $<200-250$~\kms, while shock-dominated regions can reach beyond $>200-250$~\kms. At the location of the ionized bicone in NGC~7582, we can see from the gas velocity dispersion map (Figure~\ref{fig:vel}e) that the values remain below that threshold, and thus support the interpretation of AGN-photoionized cones. However, there could be shock contributions over the South-West outer extension of the front cone, where the velocity dispersion reaches beyond 250~\kms (Figure~\ref{fig:vel}e), and the \sii/\ha\ ratios are elevated (LINER branch on Figure~\ref{fig:ratio}c; which also tend to be the same spaxels that occupy the faint right-hand LINER branch on panel a).

\section{Discussion}\label{sec:discu}

\subsection{Role of KDC for Outflow Collimation}\label{sec:colli}

Using Fabry-Perot observations, \citet{mor85} had posited the presence of a $\sim1$\,kpc \ha\ disk that was possibly rotating faster than the larger scale rotation pattern, and in any case distinct from the high-excitation gas (i.e., \oiii) kinematics. In this work, we revealed the presence of a KDC using the stellar velocity map and stellar velocity profile along the major axis (\S\ref{sec:kin}). The comparable angle and size suggest that they correspond to the same structure, which we interpreted as a rotating ring of gas, dust and stars \citep{jun20}. It also likely corresponds to the nuclear dust lanes observed with the HST by \citet{mal98}. The latter were interpreted by \citet{pri14} as the agent responsible for the collimation of the observed \oiii\ cones from narrow-band imaging \citep{rif09} given that their shape follows along the base of the broad ionization cones. We favor a similar interpretation because of geometry arguments. Namely, if we assume that the KDC is part of the same physical structure as the ring-shaped molecular gas component seen in ALMA observations by \citet{alo20,gar21}, the base of the AGN-photoionized outflows could correspond to the KDC ring opening, which is however not directly resolved with our MUSE observations. Another geometry argument comes from the misalignment between nuclear Type 1 signatures (implying that the AGN-driven outflows would be directed toward us if it is indeed a {\it true} Type 1) and the projected inclination of the cones could imply that the initial outflows partially bounce off the inner walls of the ring and get redirected to create the observed front and back cones. A consequence of either of these scenarios would be that the galaxy structure can affect the impact of AGN feedback onto the host galaxy. Collimation of ionization cones by the host galaxy interstellar medium is predicted by numerical simulations of gas-rich disk galaxies that develop a clumpy structure \citep{roo15}, supporting a potential role of galaxy substructure. 

Other observational results include the case of the Circinus galaxy, which was reported to have AGN collimation due to $\sim10$\,pc dust lane structure rather than a sub-pc scale torus \citep{mez16}. Interestingly for this object, mid-infrared observations showed that some of the emission was extended in the polar direction, which can be interpreted as dusty polar outflows \citep{stal17,stal19}. There could also be an interesting parallel with NGC~1266, whose molecular gas outflows seem to originate from a $\sim$120 pc scale nuclear gaseous disk and yet be AGN powered \citep{ala11}. Remaining sources of uncertainties to assess how galactic structure affects AGN feedback include the possible contribution from starburst-driven winds, which can be concurrent with AGN-driven winds \citep[see review by][and references therein]{vei05}, as well as multi-phase outflows including ionized gas, neutral atomic and molecular gas \citep[e.g.,][]{rup13}. In the case of NGC~7582, the outflows studied in this paper are AGN photoionized (\S~\ref{sec:ratio}), which means that they receive more radiation from the accretion disk than mechanical energy from, e.g., shocks. This feature makes it unlikely that the outflows have a starburst origin. Furthermore, the interpretation of the AGN-photoionized outflows as hollow cones is similar to that of other nearby AGN host galaxies, such as NGC~4945, which was observed with MUSE as part of the MAGNUM\footnote{Measuring Active Galactic Nuclei Under MUSE Microscope} survey \citep{ven17}, and such as HE~1353$-$1917, observed with MUSE as part of the Close AGN Reference Survey \citep[CARS;][]{hus19}. 

Similarly to other nearby barred spiral galaxies hosting low-luminosity AGN \citep{com19}, high-resolution ALMA observations of 351~GHz continuum and of CO(3-2) molecular gas in NGC~7582 revealed a distinct ring at $r>150$~pc, likely dominated by thermal dust emission, and a smaller scale nuclear disk/ring with $r<50$~pc \citep{alo20,gar21}. The latter may have an equatorial, torus-like geometry based on its alignment with respect to the outflows but more work is needed to confirm if the opening angle of this smaller structure is fully consistent with the large-scale ionized gas outflow, or if the inner outflow may still be affected by the molecular gas ring at the KDC scale. In particular, there would still need to be a potential misalignment to explain the observed Type~1 signatures and the projected angle of the collimated outflows. Even for nearby AGN hosts, there are still major open questions regarding how different types of ionized outflows (e.g., warm absorbers, broad absorption lines, ultrafast outflows) may be physically related \citep[see review by ][]{lah21}, and further connected with large-scale molecular gas outflows. We suggest that the geometry of both the nuclear and circumnuclear regions may need to be taken into account for a complete understanding of AGN-driven outflows, and their role in galaxy evolution.

\subsection{AGN Obscuration}\label{sec:torus}

There are lines of evidence showing the presence of AGN obscuration at several scales \citep[see review by][]{bia12,buc17,lah20}. This multi-scale obscuration applies to NGC~7582, which is known to be heavily absorbed in X-rays. X-ray observations obtained over the course of the previous decades have suggested the presence of at least two absorbers on different scales, including a thicker ($N_H\sim 10^{24}$\,cm$^{-2}$) absorber on small scales, and a thinner ($N_H\sim 10^{22}$\,cm$^{-2}$) absorber on large scales \citep{bia07,pic07,bia09,bra17}. The small physical scale of the heavy absorption has namely been revealed thanks to X-ray variability \citep{pic07,bia09,riv15}. These structures could correspond, respectively, to a broad line region cloud, and to large-scale host galaxy obscuration from the dust lanes. In this work, we estimated that the KDC contributes more strongly than the larger scale dust lanes. As described in Section~\ref{sec:dust}, we calculate $N_H\sim2.1\times10^{23}$~cm$^{-2}$ from the KDC. This comparatively higher column density relative to larger scale dust lanes is consistent with the extended X-ray maps from \citet{bia07}, who compared emission in a soft (0.3-0.8~keV) and slightly harder band (0.8-1.3~keV), and found an elevated hard/soft ratio behind the large-scale dust lanes suggesting $N_H\sim5\times10^{22}$~cm$^{-2}$ material is present and no detection in either band within the KDC, consistent with thicker column densities in the central $<$1~kpc.

Large-scale obscuration was also inferred from the elevated Si 9.7 absorption around $\tau_{9.7} \sim 0.7-0.8$ \citep{gou12}. In fact, this previous suspicion of AGN obscuration due - at least in part - to large-scale obscuration motivated our interest in this target in the first place. The multi-wavelength information including the optically derived obscuration from this work supports the presence of multi-scale absorbers, and differentiates between intermediate galaxy substructure scale, and large {\it full} galaxy scale. In details, there are still degeneracies in interpreting the obscuration from the infrared regime. At the $1-10$~pc ("torus") scales, there is a known degeneracy between flared disk and halo emission, which is analogous to disk orientation versus torus opening angle in AGN \citep{vin03}. Recently, \citet{bal18} modeled the NuSTAR spectra of NGC~7582 with a new set of X-ray spectral templates and inferred a line-of-sight column density to the AGN of $N_{H,los}\sim4.4\times10^{23}$~cm$^{-2}$ for an otherwise average torus column density of $N_{H,torus}\sim3.1\times10^{24}$~cm$^{-2}$ with a covering fraction of 0.9, and inclination cos($\theta_{inc}$)=0.87. According to this scenario, we could be looking into a hole through an otherwise sphere-like torus. While that modeling effort is of interest to constrain possible geometries of the absorbers, it does not constrain the physical scale. \citet{bal18} reported that their high covering fraction is consistent with results from the IR spectral energy distribution analysis by \citet{alo11}, though higher than the covering factor of $\sim0.5$ inferred by \citet{lir13}.
Future work combining IR as well as X-ray observations and models \citep[e.g.][]{,brig11b,lan19} may help clarifying the complex geometry around AGN. 

Given that our estimate for the KDC column of $N_H\sim2.1\times10^{23}$~cm$^{-2}$ is close to the line-of-sight column from \citet{bal18}, we postulate that the KDC as a galaxy substructure is a clear contributor to the average AGN obscuration. Using ALMA observations, \citet{gar21} inferred a column density of molecular gas of $N_{H_2} \sim 3\times10^{22}$~cm$^{-2}$ toward the nucleus, supporting the significant presence of gas on circumnuclear scales. However, these do not account for short episodes of higher absorption reaching the Compton-thick regime ($N_H>10^{24}$~cm$^{-2}$) mentioned above, which are likely to instead occur within the dust sublimation radius and be related to dense clumpy clouds within the broad line region \citep{pic07,bia09,riv15}. Considering the galactic kpc-scale dust lanes, sub-pc scale BLR region, and the 100~pc-scale KDC, there might be different absorber regimes contributing to AGN obscuration, with our work here emphasizing the role of the KDC.

The ALMA results mentioned earlier \citep{alo20,gar21} add yet another scale given that the KDC contains both a $r\sim$150-200~pc molecular gas ring and an inner $r<50$~pc torus-like structure. This smaller nuclear disk/ring may correspond to the structure predicted by a radiation-driven fountain model, where AGN radiation feedback induces vertical gas flows that result in a geometrically thick torus \citep{wad12}. This mechanism would lead to tori that are a few tens of pc wide, and are dynamic, evolving structures as proposed to interpret several recent observational studies and compilations \citep[e.g.,][]{ram17,hoe19,com19}. Furthermore, \citet{alo21} analyzed high-resolution mid-IR imaging of NGC~7582 from VLT/VISIR, and found both an unresolved component and an extended polar dust component. These authors argue that these observations can be interpreted with a torus+wind model, according to which IR radiation pressure creates a dusty wind component that contributes to AGN obscuration \citep[e.g.,][]{ven20}. Recent models with a realistic 3D distribution of clumpy dusty material can also reproduce polar mid-IR emission starting with a standard clumpy models, depending on the torus opening angle and scale height \citep{nik21}. Sub-parsec resolution observations will be needed to better constrain torus parameters. For example, NGC~1068 observations with the GRAVITY instrument on the European Southern Observatory Very Large Telescope Interferometer revealed a thin ring with inner radius of $r=0.24$~pc, close to the sublimation radius and inconsistent with a geometrically and optically thick torus on these small scales \citep{gravity20}. Future high-resolution observations of the nuclear regions of AGN hosts will augment our understanding, but will likely still need to be combined with probes on a range of physical scales to establish the full picture of AGN obscuration and AGN outflows.

\subsection{Comparison with Radio Continuum Emission}\label{sec:rad}

Previous work revealed extended radio continuum emission at the center of NGC~7582 \citep[e.g.][]{ulv84,for98,mor99} with two different interpretations: radio jet or starburst emission, where the latter would originate from starbursting regions along the KDC. 
Here, we will first compare the spatial extent and alignment of the 8.6~GHz radio continuum emission with the stellar and gas kinematics that we derived from the MUSE data, and then gather multi-frequency radio observations from the literature in order to estimate the age of the jet in the radio jet scenario.

\begin{figure*}
\epsscale{1.15} \plotone{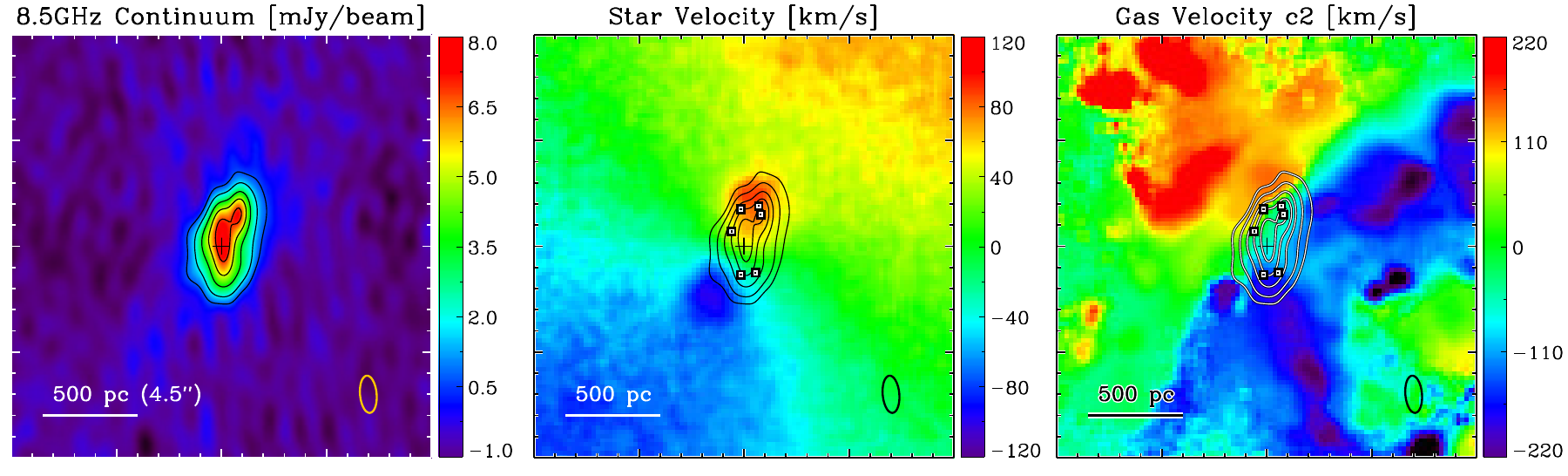}
\caption{
   (Left) Continuum at 8.6~GHz (3.5~cm) from ATCA observations of NGC~7582. The radio continuum emission likely consists of a core and diffuse components (Section~\ref{sec:rad}). Contours correspond to 1.5, 2.5, 4, 5.5, 7 {mJy\,beam$^{-1}$}. 
   (Center) Stellar velocity field as in Figure~\ref{fig:vel}, shown here on the same scale as the 8.6~GHz contours. 
   (Right) Gas velocity field for the second (outflowing) component, shown here on the same scale as the 8.6~GHz contours. The locii of six mid-infrared knots identified by \citet{wol06b} are indicated with square symbols in the central and right-hand panels.
   All panels span $20\arcsec \times 20\arcsec$ centered on the target. North is up; East is left. The scale bar measures 4\farcs5, which corresponds to 500~pc, and the ellipse shows the 8.6~GHz beam.
}\label{fig:hi}
\end{figure*}

We reprocessed the 8.6~GHz radio continuum data from \citet{mor99}.  We retrieved the data from the ATNF archive (Project C405), and obtained a final noise level of 0.175~mJy/beam with a beam FWHM of 1.763\arcsec\ by 0.792\arcsec, and a position angle of -4.9 degrees. The resulting moment zero map, and rendered beam are displayed in Figure~\ref{fig:hi}. We show the radio continuum on its own in the first panel, and superimpose the radio contours onto the MUSE stellar kinematics around the KDC, and onto the gas kinematics for the outflowing component in the middle and last panel of Figure~\ref{fig:hi}. From the comparison with the stellar kinematics, we can see that the extended radio emission is generally aligned and overlaps with the North side of the KDC, and with a possible extension south of the nucleus, which could be slightly offset from the South side of the KDC. Based on 3.5~cm and 6~cm radio maps, \citet{for98} found that the South component, which they presume is associated with the true nucleus, has a spectral index of $\alpha=-0.7$, consistent with non-thermal emission from an AGN. The northern component has a spectral index of $\alpha=-0.9$, which could possibly be associated with a starburst origin.

Given that both the radio emission, and the KDC morphology are only marginally resolved, we lack information to confirm whether the radio emission is a short jet trapped within $-$ or interacting with $-$ the KDC, or whether it arises directly from the starbursting activity embedded in the KDC. We note that the radio continuum is not seen over the South portion of the KDC (blue in the MUSE velocity map), which suggests one of these options: (i) the starburst is unevenly distributed around the circumnuclear region, (ii) extended radio emission was resolved out given the configuration of the antennas, (iii) the radio jet scenario is a viable explanation, perhaps similar to the cases reported by citet{ven21}, where compact, low-inclination radio jets propagate through the host galaxy disks and cause extended ionized gas outflows.

\citet{ricc18} conducted a detailed study of the central region of NGC~7582 to investigate AGN, stellar, and/or shock emission by combining various optical and infrared observations. Namely, these authors presented spatially resolved flux maps showing two blobs of emission seen in 
\feiilam, \htwo\ and \brgam, one North and one South of the nucleus. They found that these blobs of emission coincide with previously identified knots of mid-infrared emission \citep[][]{wol06b}, and were aligned with the radio continuum data from \citet{mor99}. However, they also found interesting differences between the two blobs in the sense that the South blob overlaps with three MIR knots \citep[and with emission seen in a narrow-band filter at 11.88$\mu$m spanning PAH features][]{asm14}, and is likely dominated by shocked gas, whereas the North blob, which overlaps with the other three MIR knots, is likely dominated by photo-ionized gas. In particular, \citet{ricc18} reported that the two West-most MIR knots (squares on Figure~\ref{fig:hi}) have emission-line signatures indicating partially-ionized zones such as elevated \oilam, and may lie along the edges of the front ionized cone.
Based on their analysis of gas excitation conditions and geometry of knots of emission in the circumnuclear region, \citet{ricc18} favored a scenario with shock excitation due to an impact from radio jets onto the circumnuclear medium, and proposed a geometry with a radio jet strongly inclined with respect to the accretion disk/torus (their Figure 17).
If these knots of starburst activity were indeed triggered by the impact from radio jets or AGN-driven outflows, we may be witnessing a case of positive AGN feedback such as reported in previous studies of, e.g., NGC~5643 \citep{cre15}, and as predicted from numerical simulations of a radio jet traveling through a clumpy disk \citep[e.g.,][]{gai11}. 

Regarding the gas kinematics, the radio emission appears to overlap spatially with the base of the ionized gas outflows (Figure~\ref{fig:hi}) but with a tilt relative to the direction of the outflowing gas. This behavior resembles the predicted outcome of radio jets propagating through clumpy gaseous disks, where outflows end up following the path of least resistance and traveling preferentially perpendicular to the disk even when the jet is nearly aligned with the disk plane \citep{muk18,muk18b}. Tentatively, our interpretation of the KDC as a ring could be consistent with a geometry involving a strongly tilted jet, which in turns carves out the central part of the KDC (i.e., turning a disk into a ring), and/or that are bouncing off the edges of that structure, similarly to what we suggested for the \oiii\ outflows in Section~\ref{sec:colli}. This scenario could explain the tentative deviation of the Northern extension of the radio emission where it may collide with the KDC. However, the current spatial resolution is insufficient to confirm this hypothesis. In order to distinguish between the potential scenarios explaining the nature of the extended radio emission, one would need high resolution observations with at least two different frequencies to measure the radio spectral index of both the point source and the extended components.

If both the radio jets and ionized gas outflows were re-collimated by the KDC, the physical implication would be that the presence of galaxy substructure on circumnuclear scales can influence their geometry, and therefore affect the impact that AGN feedback from radio jet and/or accretion disk winds can have on host galaxies. Indeed, by redirecting the AGN feedback preferentially perpendicular to the galactic disk, the central substructure effectively reduces the direct impact of AGN-driven outflows. Another implication is that low-power radio jets may contribute to drive outflows even in radio quiet galaxies as suggested by, e.g., \citet{hus19} from their study of edge-on disc galaxy HE 1353-1917, which shows AGN ionization cones aligned close to the galaxy plane out to $\sim$25~kpc scales, and including AGN driven outflows on $\sim$1~kpc scale in alignment with the low-power radio jet. More recently, \citet{ven21} reported findings of large-scale outflows within four nearby galaxies with compact, low-power radio jets which shared the characteristic of having a low inclination with respect to the galaxy disk. These authors suggest that the low inclination enables more significant AGN feedback by increasing the amount of ISM material impacted by the jets as they propagate. From another case study, \citet{garb21} reported that both radio jets and AGN disk winds may act simultaneously when considering multiphase (molecular and ionized gas) outflows in Sy2 galaxy NGC~5643.

Lastly, we add another piece to the NGC~7582 puzzle by estimating the age of the radio jet under the hypothesis that the extended emission corresponds to a low-power jet. To do so, we combine multi-frequency radio continuum measurements. Namely, \citet{ori10} analyzed VLA data for NGC 7582 at 4.8 and 8.4~GHz, reporting flux densities of 75~mJy, and 42~mJy, respectively.  They further separated the core radio emission from the diffuse star forming radio emission to determine that the core has a spectral index of 0.6 and is unresolved at less then 40~pc in linear size.  They also calculated the equipartition magnetic field to be around 1.5$-$10~mG for their sample of Seyferts.  At shorter frequency, \citet{mau03} measured 395.3~mJy at 0.843~GHz, and \citet{con96} reported 262~mJy at 1.4~GHz. NGC~7582 was not detected in the Australia Telescope 20 GHz Survey \citep[AT20G;][]{mur10}. Based on these literature values, the break frequency needs to be between 5$-$20~GHz. We used this range to bracket the corresponding age of the radio AGN assuming the continuous injection model from \citet{mur99}. Considering no expansion and assuming that the magnetic field remains constant then the elapsed time since source formation is given by their equation 2, and places the age of the AGN around 11,000 to 23,000 years old. This young age is consistent with the projected length of the jet. To cover a distance of 300~pc, the jet would need to travel at 4-10\% the speed of light during the elapsed time. While these numbers are plausible, current observations are not sufficient to confirm the nature of the extended radio emission.

\subsection{Possible Origin of the KDC}\label{sec:kdc}

KDCs have been observed in the center of several galaxies, and exhibit various sizes and properties, including counter-rotating, co-rotating, or non-rotating cores \citep[e.g., ][]{ems04,kra08,rai13}. There are several examples from studies of early type galaxies such as from the SAURON \citep{dez02}, and ATLAS$^{3D}$ \citep{capp11} surveys. The physical origin of such distinct kinematics in the cores of galaxies is usually associated with the effect of a tidal interaction or with the resulting perturbation of the galaxy kinematics \citep[e.g.,][]{her91}. Using an analysis of the full SAURON sample, \citet{mcd06} found indications suggesting two predominant categories of KDCs: large kpc-size KDCs with old stellar populations ($>8$~Gyr) that reside in so-called {\it slow rotators} \citep{ems07}, and compact KDCs (few hundred pc) with a broad range of stellar ages but found primarily in {\it fast rotators}. Detailed studies of the lenticular (S0) galaxy MCG-06-30-15 \citep{rai13,rai17} revealed a $\sim125$~pc scale counter-rotating KDC in stellar kinematics, which was further shown to be co-located with a counter-rotating disk of warm molecular gas traced by $H_2$. The co-location of the stellar and warm molecular gas components sharing similar kinematics is consistent with a scenario where a KDC can include both stars and gas in the same physical structure, as we posit is also the case for NGC~7582. Using MUSE, \citet{joh18} presented the analysis of elliptical galaxy NGC~1407 and found that it contains a KDC on a comparable physical scale of $\sim0.6$~kpc as what we find for NGC~7582. These authors attribute the cause of the KDC to either a major merger or a series of minor mergers.

Large- and small-scale KDCs may originate from different types of gravitational interactions with other galaxies or instabilities \citep[e.g.,][]{boi11}. In the case of barred spiral galaxies, the presence of the bar can induce rings at the location of the inner Lindblad resonance \citep[ILR;][]{lin64,lin74}. In their review, \citet{but96} compiled both observational and theoretical investigations of rings in galaxies, including findings that rings are most typically associated with bars (or ovals, another common nonaxisymmetric perturbation). The bar formation itself is likely due to previous or ongoing gravitational interactions with a companion galaxy, or even a close passage \citep{ger90,pet18}. Bar formation can also result from internal dynamical instabilities, which are predicted to lead to different bar properties \citep[e.g.][]{mar17}.

NGC~7582 is a member of the Grus Quartet, a group of four spiral galaxies that have been observed to be experiencing some gravitational interactions based on tidal features seen in \hi\ emission \citep{kor96b,dah05,fre09}, and \hi\ absorption \citep{kor96}. In particular, \citet{kor96b} identified a long tail of \hi\ gas extending from NGC~7582. \citet{dah05} computed that the tail comprises $\sim1.3 \times 10^9$\,\Msun\ of gas. They also found a large \hi\ cloud of $7.7 \times 10^8$\,\Msun\ that was likely expelled from NGC~7582 into the intergalactic medium. The other large spirals do not show a similar tail, though \citet{dah05} reported that NGC~7552 has a much shorter \hi\ extension (or tail) pointing away from the rest of the group.  Therefore, the long \hi\ tail from NGC~7582 can be attributed to either a minor merger, where a smaller dwarf companion lost its reservoir of \hi\ gas creating the \hi\ extension on its way to merging with NGC~7582, or possibly to a minor interaction with the distant NGC~7552 quartet member, though this seems less likely given that the NGC~7552 \hi\ feature is much weaker. In either scenario, such a disturbance could have led to the presence of the large-scale bar in NGC~7582.

Putting the pieces together, we interpret that the KDC is possibly at the ILR of the bar, and that the bar may itself have been created during a previous interaction with a small companion that interacted with the Grus quartet and in particular with NGC~7582. However, we note that other members of the quartet have been reported to have bars and that NGC~7552 was reported to also harbor a circumnuclear starburst ring \citep{for94}. Therefore, it is possible that the bars result from some previous or ongoing gravitational interaction between NGC~7582 and NGC~7552. Lastly, we also considered a possible origin from a radio jet propagating within the KDC in Section~\ref{sec:rad}. 

Other studies of interest include the use of Fabry-Perot observations to fully map the dynamics of nearby barred galaxies such as  NGC~4123 \citep{wei01}, NGC~1433, NGC~6300 \citep{but01}, and NGC~1365 \citep{zan08} which all have a bar and rings and/or pseudorings. Gas streaming along the bar potential can feed the ILR ring, which in turn can become a reservoir for smaller gas structure. This was posited by \citet{com19} as they further found indications that some ILR rings have gas streams or trailing spiral arms linking them to small-scale molecular tori, consistent with theoretical expectations for inflows from the ring toward the central region. Another possible explanation for the KDC is a nuclear stellar disk that grows inside-out as the bar evolves and creates successive ILR rings with increasingly larger radii, such that the current ILR ring co-exists with the nuclear disk, at its outer edge \citep{gad20,bit20}.

Additional work such as detailed modeling of the stellar and gas kinematics may shed more light on the physical origin of the KDC, noting that the KDC was observed in both the stellar and gas velocity maps (Figure~\ref{fig:velpro}), and that the molecular gas ring revealed by recent ALMA observations was also attributed to the ILR \citep{alo20,gar21}.

\section{Summary}\label{summ}

In this paper, we presented VLT/MUSE observations obtained in order to investigate the kinematics and ionization properties of nearby galaxy NGC 7582, which harbors an extremely obscured AGN. We searched for physical links between the central AGN and host galaxy at various scales including the role of galactic substructure in changing the impact of AGN feedback and contributing to AGN obscuration.  Our main findings are the following:

\begin{enumerate}

\item[1.] The large-scale stellar and nebular gas velocity fields reveal the presence of large-scale disk rotation, a kinematically distinct core (KDC), and conical bipolar outflows. The kinematics of the stellar KDC is reported here for the first time, and is attributed to a $\sim$600~pc diameter ring (or disk) of stars, gas and dust. The main gas velocity field largely follows the same large scale rotation as the stellar velocity field but with a systematic offset and hints of kinematics affected by the presence of the previously-known large-scale galactic bar.

\item[2.] Emission line ratio maps show that the bipolar cones are primarily photoionized by the AGN, with Seyfert-like signatures. Regions on either side of the cones may instead be shock-ionized, with LINER-like line ratios. Lastly, gas along the main galactic disk is shaped in star-forming knots or clumps with emission line ratios consistent with stellar photoionization. We detect higher excitation in diffuse regions of the galaxy disk, perhaps due to leaking AGN radiation through the disk, or otherwise corresponding to low density WHIM.

\item[3.] The morphology and kinematics of the ionized cones are consistent with the collimation taking place at the scale of the 600~pc ring. From the literature, the accretion disk must be oriented with a low inclination angle, while the ring and cones have a significant inclination. Thus, we postulate that the ring might also deflect AGN-driven winds. If this scenario is true, a broader implication is that galaxy substructure can play a role in how AGN feedback affects galaxies.

\item[4.] Dust obscuration maps show a peak along the central dust lanes previously observed by \citet{mal98} for both stellar and gas attenuation, and consistent with the analysis presented by \citet{pri14}. The stellar-to-gas attenuation ratio varies from values similar to that of Calzetti (0.44) in star-forming regions in the disk, reaching higher than 0.44 in diffuse regions, and much lower in the North-West portion of the disk indicating heavily obscured \hii\ regions, which may be related to a previous or ongoing minor merger.

\item[5.] Contributions to AGN obscuration take place at multiple scales: small (sub-torus) scale variable X-ray absorption that varies between Compton-thick and Compton-thin \citep[$N_H>10^{23-24}$cm$^{-2}$;][]{pic07,bia09,riv15}; the KDC may contribute to the $10^{23}$ level; galaxy-scale dust lanes are estimated to have column densities $N_H\sim10^{22}$ \citep{bia09}. While the role of galactic substructure such as the KDC may not dominate compared to small sub-pc scales, it has the potential to be much more important than galaxy-scale dust lanes.

\item[6.] The physical origin of the KDC is possibly due to an inner Lindblad resonance with the bar. The bar itself could have formed from a minor merger or a gravitational interaction with other massive galaxies that are part of the Grus Quartet. Alternatively, the KDC might be connected to the presence of a radio jet or a direct consequence of a minor merger. Future work involving kinematic modeling will help answer this question.

\end{enumerate}

This case study of NGC~7582 revealed that galaxy substructure can play intriguing roles in shaping the connection between AGN and their host galaxies. Analysis of previous and upcoming observations using integral-field-spectroscopy, such as CARS \citep{hus19}, TIMER \citep{gad19}, KOALA \citep{u19}, AMUSING++ \citep{lop20} and others are a promising tool to reveal a more complete picture. In particular, transitioning from case studies to larger, ensemble studies will give us a more systematic view of the impact of AGN feedback in affecting host galaxies via radiative and/or kinetic feedback. Learning from this work, we stress the importance to achieve sufficient spatial resolution to probe galactic (sub)structure on physical scales of tens to a few hundred parsecs.

\acknowledgments 

This paper is dedicated to the memory of Michael Dopita. Mike contributed to this project with unmatched enthusiasm from the very beginning, and suggested significant improvements to the analysis, and to this manuscript. We warmly thank the anonymous referee for suggestions which greatly augmented this work, in particular for prompting a more careful analysis of the KDC velocity and velocity dispersion profiles. 
SJ thanks J. Najita, K. Olsen, E. Walla, S. Ridgway, F. Combes, J. Silk, F. Schweizer, D. Rupke, M. den Brok, E. Momjian, L. Fortson, and G. Bicknell for useful and enlightening discussions.

Based on observations made with ESO telescopes at the La Silla Paranal Observatory under programme ID 095.B-0934. Some of this research uses services provided by the Astro Data Lab at NSF's National Optical-Infrared Astronomy Research Laboratory. NOIRLab is operated by the Association of Universities for Research in Astronomy (AURA), Inc. under a cooperative agreement with the National Science Foundation. 
Parts of this research were supported by the Australian Research Council Centre of Excellence for All Sky Astrophysics in 3 Dimensions (ASTRO 3D), through project number CE170100013, and by the Australian Research Council Centre of Excellence for All-sky Astrophysics (CAASTRO), through project number CE110001020. SB acknowledges financial support from the Italian Space Agency under grant ASI-INAF 2017-14-H.O. 
FEB acknowledges support from ANID-Chile BASAL AFB-170002 and FB210003, FONDECYT Regular 1200495 and 1190818, 
and Millennium Science Initiative Program ICN12\_009.
JS is supported by the international Gemini Observatory, a program of NSF's NOIRLab, which is managed by the Association of Universities for Research in Astronomy (AURA) under a cooperative agreement with the National Science Foundation, on behalf of the Gemini partnership of Argentina, Brazil, Canada, Chile, the Republic of Korea, and the United States of America.
JHW acknowledges the support by the National Research Foundation of Korea grant funded by the Korea government (NRF-2021R1A2C3008486).

\facilities{VLT:Yepun, ATCA, Astro Data Lab}

\software{LZIFU \citep{ho16}, MPDAF: MUSE Python Data Analysis Framework \citep{bac16} }

\clearpage

\appendix

\section{Animated Views of the NGC 7582 Spectral Cube}\label{app:anim}

\begin{figure*}[b]
\epsscale{1.} \plotone{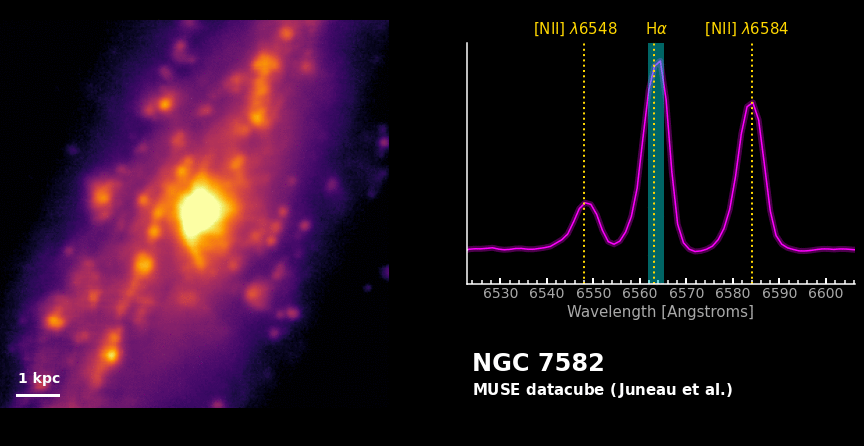}
\caption{
   Snapshot from an animation stepping through the MUSE datacube. (Left) Slice of the datacube averaged over the 4\,\AA\ window highlighted in the right panel. North is up; East is left. The scale bar measures 9\arcsec, which corresponds to 1~kpc. 
   (Right) Spectrum coadded over the MUSE field-of-view, with emission lines marked with vertical dotted lines and labeled. This animation encompasses the \ha\ and \niilamboth\ emission lines. In the animated version, the 4\,\AA\ starts from the blue end of the spectral range, and gradually shifts from blue to red wavelengths as the animation progresses, with both panels updating simultaneously. 
  This figure is available as an animation in the online journal.
}\label{fig:animHa}
\end{figure*}

\begin{figure*}
\epsscale{1.} \plotone{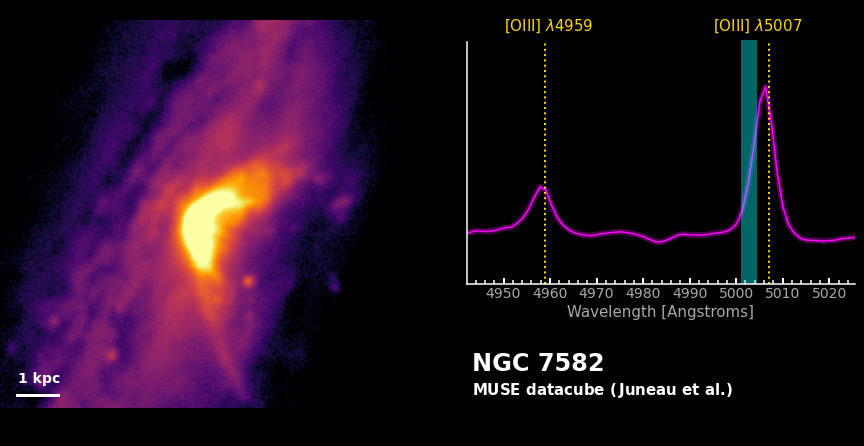}
\caption{
   Snapshot from an animation stepping through the MUSE datacube. (Left) Slice of the datacube averaged over the 4\,\AA\ window highlighted in the right panel. North is up; East is left. The scale bar measures 9\arcsec, which corresponds to 1~kpc. 
   (Right) Spectrum coadded over the MUSE field-of-view, with emission lines marked with vertical dotted lines and labeled. This animation encompasses the \oiiilamboth\ doublet. In the animated version, the 4\,\AA\ starts from the blue end of the spectral range, and gradually shifts from blue to red wavelengths as the animation progresses, with both panels updating simultaneously. 
  This figure is available as an animation in the online journal.
}\label{fig:animOiii}
\end{figure*}

To visually interpret the ionized gas kinematics and spatial distribution, we constructed animated figures that step through wavelength slices of the MUSE datacube over two spectral regions of interest. Firstly, Figure~\ref{fig:animHa} spans rest-frame wavelengths 6523$-$6606\,\AA, encompassing \ha\ and the \niilamboth\ doublet. Secondly, Figure~\ref{fig:animOiii} spans rest-frame wavelengths 4942$-$5025\,\AA, which contain the \oiiilamboth\ doublet primarily tracing AGN-ionized gas. The datacube includes stellar continuum and emission lines. We apply a spatial smoothing using a 2D Gaussian kernel with a $3\times3$~spaxel width on each individual spectral channel of the cube. To produce an animation, we calculate the mean emission in slices of 4\,\AA, and shift the spectral coverage by +1\,\AA\ for each new animation frame, producing a running mean image in thin spectral slices. The shifting wavelength window is displayed simultaneously with the corresponding image as a vertical band over the total MUSE spectrum. The latter was obtained from the sum of MUSE datacube along the wavelength direction. 

These figures add a dynamic view of some of the results presented in the main paper. The stellar continuum is approximately flat for a given animation, creating a constant background image with obvious dust lanes. On the contrary, the emission lines vary strongly both spatially and spectrally. As the animation progresses from shorter/bluer wavelengths toward longer/redder wavelengths, the window crosses each emission feature from the approaching side to the receding side. The \ha\ and \nii\ emission lines (Figure~\ref{fig:animHa}) are produced in star-forming regions in the galactic disk+bar, as well as in the ionized cone. The animation clearly shows how the star-forming regions follow a large-scale rotation pattern from lower left (South-East) toward top right (North-West), which was represented with gas velocity maps in Figures~\ref{fig:vel} and \ref{fig:velweight}. The ionized outflow is most prominent at the extreme end of the blue wing of each emission line profile, due to the outflowing gas reaching faster approaching velocities relative to the disk in the MUSE field-of-view (Figure~\ref{fig:vel}b,c). Faint but large-scale extensions to the ionized cones are also noticeable in the blue wing of the \niilam\ line profile, reaching over 3~kpc from the center (in projection).

The animation around the \oiii\ doublet (Figure~\ref{fig:animOiii}) also includes the stellar continuum. In contrast with the previous animation, the ionized gas is mostly located in the outflowing cones. The blue wing of each emission line originates from gas that is outflowing toward our line of sight. In the case of NGC~7582, this component of the outflow dominates the \oiii\ doublet emission, causing the peaks to be slightly blueshifted with respect to the expected rest-frame wavelengths (vertical dotted lines). It is interesting to note that some of the outflowing gas which is receding (redshifted), is visible both along the front cone (possibly from the far side of the front cone) and along the back countercone. As we noted in the manuscript, the latter is mostly visible between the dust lanes, indicating that the back cone is located behind the galaxy disk (Figure~\ref{fig:labels}). The animations were produced using the MUSE Python Data Analysis Framework \citep[MPDAF, ][]{bac16} package and Matplotlib Python library on the Astro Data Lab science platform \citep{nik20,jun21}.

\clearpage

\bibliographystyle{aasjournal}
\bibliography{bib_agn}

\end{document}